\documentclass[a4paper,11pt]{article}
\pdfoutput=1 

\usepackage{jcappub} 

\usepackage[T1]{fontenc} 
\usepackage {amssymb}
\usepackage {amsmath}
\usepackage{mathabx}
\usepackage{float}

\usepackage{xcolor}

\title{\boldmath Towards understanding thermal history of the Universe through direct and indirect detection of dark matter}

\author[a,b]{Leszek Roszkowski,}
\author[a,c]{Sebastian Trojanowski}
\author[d]{and Krzysztof Turzy\'nski}

\affiliation[a]{National Centre for Nuclear Research,\\Ho{\. z}a 69, 00-681 Warsaw, Poland}
\affiliation[b]{Department of Physics and Astronomy, University of Sheffield,\\Sheffield S3 7RH, United Kingdom}
\affiliation[c]{Department of Physics and Astronomy, University of California, Irvine,\\California 92697, USA}
\affiliation[d]{Institute of Theoretical Physics, Faculty of Physics, University of Warsaw,\\ul.~Pasteura 5, 02-093, Warsaw, Poland}

\emailAdd{leszek.roszkowski@ncbj.gov.pl}
\emailAdd{sebastian.trojanowski@uci.edu}
\emailAdd{Krzysztof-Jan.Turzynski@fuw.edu.pl}

\abstract{We examine the question to what extent prospective detection of dark matter by direct and indirect- detection experiments could shed light on what fraction of dark matter was generated thermally via the freeze-out process in the early Universe. By simulating putative signals that could be seen in the near future and using them to reconstruct WIMP dark matter properties, we show that, in a model- independent approach this could only be achieved in a thin sliver of the parameter space. However, with additional theoretical input the hypothesis about the thermal freeze-out as the dominant mechanism for generating dark matter can potentially be verified. We illustrate this with two examples: an effective field theory of dark matter with a vector messenger and a higgsino or wino dark matter within the MSSM.
}


\arxivnumber{1703.00841}

\begin{document}
\maketitle
\flushbottom

\section{Introduction}

A detection of non-gravitational interactions of 
dark matter (DM) particles 
would be one of the biggest breakthroughs in contemporary physics.
There are many
experimental strategies to search for such a signal,
which include direct detection (DD) of DM particles by their scattering off heavy nuclei in deep underground detectors, indirect detection (ID) of products of DM annihilations or decays with space and ground based telescopes, and collider searches. 
So far, these efforts have led to stringent limits on DM interactions, but many of the favorable and very promising theoretical scenarios have not been fully probed yet, \textsl{e.g.}, the prototypical scenario of a weakly interacting massive particle (WIMP), which we will denote by $\chi$, such as the lightest neutralino in the Minimal Supersymmetric Standard Model (MSSM) (for review see, \textsl{e.g.}, \cite{Martin:1997ns}).

Gravitational interactions of DM point towards a very precise value of its contribution to the energy density of the Universe, $\Omega_\chi h^2=0.120\pm0.003$
 \cite{Ade:2015xua}. 
 It is most often 
 considered to have arisen
in the so-called thermal production of DM in a freeze-out process: the decoupling of the nonrelativistic DM particles that previously were in kinetic and chemical equilibrium with the thermal plasma, see, \textsl{e.g.}, \cite{Kolb:1990vq}, and introduces a strict correlation between the mass and the interaction strength of DM particles.
One may, however, also consider DM candidates that do not fit to the standard thermal production mechanism, but can still be effectively produced in other ways (for a review see, \textsl{e.g.}, \cite{Baer:2014eja}).
If the DM particles couple too strongly to the SM sector, 
the DM abundance is reduced and one needs to consider an additional mechanism of DM production.

In this paper, we aim at studying the properties of DM particles in a new way with the ultimate goal of understanding better the thermal history of the early Universe. 
We adopt a view that within the next decade or so collider searches for physics beyond the Standard Model will give negative results, but there will be
a WIMP DM detection 
in one or more of the following experiments:
Xenon1T~\cite{Aprile:2015uzo} direct search for DM particles, the FermiLAT search for DM induced $\gamma$ rays \cite{Atwood:2009ez} and the Cherenkov Telescope Array (CTA) \cite{Consortium:2010bc}, a ground-based telescope which will also be looking for $\gamma$-ray signal from DM annihilation and is expected to begin operating in some years from now. 
However, we are going to analyze the prospective signal cautiously and conservatively, taking into account known uncertainties. One can envisage a few possible scenarios.
Given the sensitivities of the experiments, it is quite plausible that a positive DM signal would
correspond to interactions strength larger than required for thermal DM production, which would point towards the existence of 
another DM component, presumably the particles of the same identity but originating from a process different from freeze-out. 
Another goal of ours is to characterize, to the largest possible extent, that additional component and the mechanisms that gave rise to it.
To this end, we simulate the signals that can be obtained by the collaborations in the case of a discovery and, subsequently, we use them to derive the basic properties of the DM particles, \textsl{e.g.}, $m_\chi$ and the today's value of the thermal average of the annihilation cross section times the M\o ller velocity (from now on referred to as annihilation cross section), $\langle\sigma v\rangle_0$, with their respective uncertainties, including astrophysical interpretation (for previous similar studies for the Fermi-LAT see \cite{Bernal:2010ip,Bernal:2011pz}).
In our approach it is possible to discuss these constraints at various levels of theoretical assumptions: from a model-independent level in which it is enough to introduce a set of phenomenological parameters to describe the relevant experimental signatures that we take into account, through an effective field theory of DM to very specific MSSM examples.

To this end, in Section \ref{sec:gen} we perform a model-independent analysis and conclude that, given the current limits on $\langle\sigma v\rangle_0$ and known uncertainties, it would be very difficult, if not impossible, to rule out the possibility that DM consists only of the freeze-out component, even in the simplest models, not to mention scenarios of multi-component dark matter or with Sommerfeld enhancement (SE). In Section \ref{sec:eft}, we move on to describing DM within the framework of an effective field theory of a vector messenger, focusing on additional information that can be obtained from including positive signals from direct detection experiments. In Section \ref{sec:neutralino}, we analyze the specific case of the MSSM with neutralino DM. In this framework, we can discuss the mechanism for generating a non-thermal component of DM from decays of heavier particles such as a gravitino or an axino that belong to the category of extremely weakly interacting massive particles (EWIMPs). Since such particles were thermally generated at earlier stages of the evolution of the Universe, a determination of non-thermal components constrains the value of the reheating temperature. We conclude in Section \ref{sec:conclusions}. Technical details, in particular those regarding reconstruction of DM properties from the signals from direct and indirect detection experiments, are deferred to the Appendices.

\section{General WIMP DM}
\label{sec:gen}

We assume that WIMP DM could be produced in the early Universe both in freeze-out, as well as in non-thermal processes, \textsl{e.g.}, late-time decays of some heavier species. The WIMP DM relic density can then be schematically written
\begin{equation}
\Omega_\chi h^2 = \Omega_\chi^{\textrm{fo}}h^2 + \Omega_\chi^{\textrm{non-th}}h^2 
\label{eq:Oh2contributions}
\end{equation}
where $\Omega_\chi^{\textrm{fo}}h^2$ and $\Omega_\chi^{\textrm{non-th}}h^2$ are the contributions to the WIMP DM relic density from the freeze-out and non-thermal processes, respectively.

The first term on the r.h.s.\ of Eq.~(\ref{eq:Oh2contributions}) depends on 
the annihilation cross section of DM particles 
which can be often approximated as
\begin{equation}
\langle\sigma v\rangle = \frac{\alpha_s + (T/m_\chi)\,\alpha_p}{m_\chi^2} \, ,
\label{eq:sigmavWIMP}
\end{equation}
with coefficients $\alpha_s$ and $\alpha_p$ parameterizing the $s$-wave and $p$-wave contributions to $\langle\sigma v\rangle$, respectively. 
Eq.~(\ref{eq:sigmavWIMP}) can be used to relate the value of $\langle\sigma v\rangle_0$ today which can be determined from the indirect detection of DM with the value of the annihilation cross section at the DM freeze-out, $\langle\sigma v\rangle_{\textrm{fo}}$, which in turn determines the value of $\Omega_\chi^{\textrm{fo}}h^2$. 
One can then make inferences about the limits on the 
additional DM component,
presumably produced after thermal freeze-out.

While the simple relation in Eq.~(\ref{eq:sigmavWIMP}) is not always valid, \textsl{e.g.}, when coannihilations between the DM particles and other species are important around the DM freeze-out \cite{Griest:1990kh,Gondolo:1990dk}, one can still apply it for a conservative estimate, because 
additional interactions typically lead to an increase of the effective annihilation cross section at freeze-out.
There are, however, some scenarios in which these estimates do not apply. One possibility is that $\langle\sigma v\rangle_\mathrm{fo}$ is smaller than $\langle\sigma v\rangle_0$ due to interactions with other particles, \textsl{e.g.},~resonant annihilation or Sommerfeld enhancement. One can also envision a more complicated DM sector, consisting of two or more stable particles that have undergone freeze-out separately. Therefore, in the following we will first discuss simple scenarios in which (\ref{eq:sigmavWIMP}) can be applied and then turn to a few representative models with the present-day annihilation cross-section larger than at freeze-out. We also note that
within a specific DM model, such as neutralino DM with a gravitino or axino companion in the MSSM investigated in Section~\ref{sec:neutralinoDM},  one does not need to resort to the approximation~(\ref{eq:sigmavWIMP}), as a precise relation between $\langle\sigma v\rangle_0$ and $\langle\sigma v\rangle_{\textrm{fo}}$ is known.

%
%

\subsection{The simplest case: single-component DM with present-day annihilation cross-section not larger than at freeze-out}
\label{ssec:gen_sim}

The parameter space that we scan over for a model-independent study is defined in terms of the WIMP DM mass $m_\chi$, its ID cross section, $\langle\sigma v\rangle_0$, as well as the branching ratios that describe the final states for DM annihilation as shown in Table~\ref{tab:modelindepparams}. We allow DM to annihilate into several final states or their mixtures. In particular we focus on $b\bar{b}$, $W^+W^-$, $\tau^+\tau^-$ and $hh$ final states that correspond to distinct annihilation spectra. There are many other Standard Model final states, but usually they can be accurately approximated by one of the final states above or by the combinations thereof. However, we do not discuss here the purely leptonic final states, \textsl{e.g.}, $\mu^+\mu^-$, since they are typically easily distinguishable from other cases that we take into account if $\langle\sigma v\rangle_0$ is large enough. For a more detailed discussion about the reconstruction for different final states, see \cite{Roszkowski:2016bhs}.

In order to remain as model independent as is possible, in this Section we do not assume any cross correlation between $\langle\sigma v\rangle_0$ and $\sigma_p^{\textrm{SI}}$. In practice, this is equivalent to excluding DD entirely from our considerations, because of the 
well-known $\sigma_p^{\textrm{SI}}/m_\chi$ degeneracy in the recoil spectra for $m_\chi\gtrsim 100$~GeV. 
This is particularly relevant for our analysis, because it is just the mass range obtained by
combining our assumption that $\langle\sigma v\rangle_\mathrm{fo}$ is larger than the value required for thermal WIMP DM density with the 
current limits from null results of DM searches in dwarf spheroidal galaxies (dSphs) \cite{Ahnen:2016qkx,Fermi-LAT:2016uux} (see also \cite{Profumo:2016idl} for the limits relevant for monochromatic-like spectra).
In other words, irrespectively of whether DM DD signal is detected or not, for a given $m_\chi$ one can always adjust $\sigma_p^{\textrm{SI}}$ to account for
that part of experimental input.

\renewcommand{\arraystretch}{1.3}
\begin{table}[t]
   \centering\footnotesize
   \begin{tabular}{|c|c|c|c|} 
      \hline
      \textbf{Symbol} & \textbf{Parameter} & \textbf{Scan range} & \textbf{Prior distribution} \\
      \hline
      \hline
       $m_\chi$ & WIMP mass & $10-10000\,\textrm{GeV}$ & log \\
       \hline
       $\langle\sigma v\rangle_0$ & Annihilation cross section & $10^{-30}-10^{-21}\textrm{ cm}^3/\textrm{s}$ & log \\
        & (indirect detection) &  &  \\
       \hline
       $f_{i}$ & Branching ratios for various final states & $0-1$ & modified Dirichlet distribution \\
        & $b\bar{b}, W^+W^-, \tau^+\tau^-, hh$ &  &  (branching ratios add to~1) \\
       \hline
   \end{tabular}
   

   \caption{Input parameters for the model-independent reconstruction of DM properties. The nuisance parameters varied in our scans are shown in Table~\ref{tab:nuiparams}.}
   \label{tab:modelindepparams}
\end{table}

\begin{figure}[t]
\begin{center}
\includegraphics*[width=0.49\textwidth]{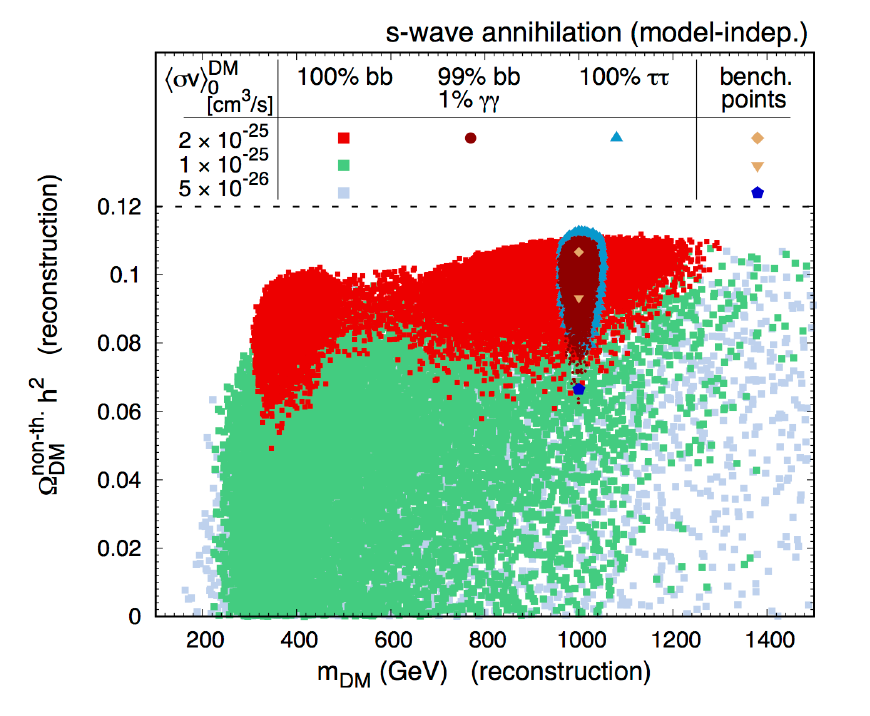}
\hfill
\includegraphics*[width=0.49\textwidth]{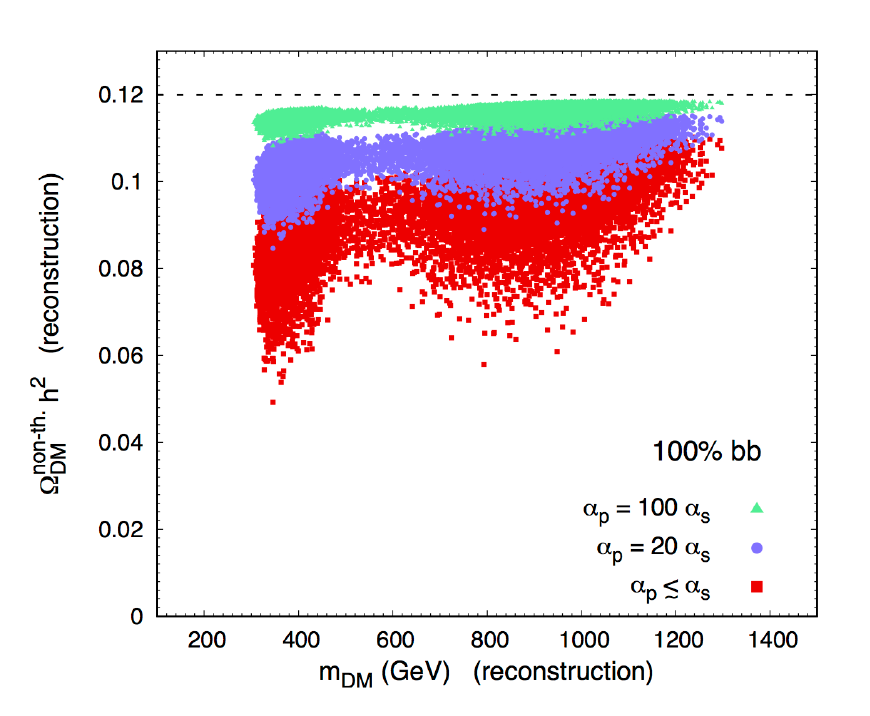}
\end{center}
\caption{Reconstruction of the non-thermal contribution to the DM relic density, $\Omega_\chi^{\textrm{non-th}}h^2$, as a function of the reconstructed mass of DM, $m_\chi$, for the benchmark point (marked by light brown diamond in the left panel) corresponding to $m_\chi = 1$ TeV. \textsl{Left panel}: The results obtained for pure $s$-wave annihilation into $b\bar{b}$ final state with and without $1\%$ admixture of the $\gamma\gamma$ channel, as well as the pure $\tau^+\tau^-$, for three values of $\langle\sigma v\rangle_0 = (2,\,1\,\,\textrm{and}\,\,0.5) \times 10^{-25}\,\textrm{cm}^3/\textrm{s}$. \textsl{Right panel}: The results obtained for pure $b\bar{b}$ final state with significant $p$-wave contribution to the annihilation cross section around freeze-out
with $\alpha_p/\alpha_s = 20$ (blue circles) and $100$ (green triangles). The pure $s$-wave case (light red squares) is also shown for comparison. In both panels the horizontal black dashed line corresponds to the total DM relic density.}
\label{fig:Oh2nonth}
\end{figure}

Once we determine the $95\%$ CL intervals for the quantities describing the would-be-detected WIMP DM particles, \textsl{i.e.}, $m_\chi$ and $\langle\sigma v\rangle_0$, we can estimate the 
contribution to the DM density originating from freeze-out,
using Eq.~(\ref{eq:sigmavWIMP}). 
The most conservative estimate comes the assumption of $s$-wave dominance, 
$\alpha_p \to 0$, 
which translates to
$\langle\sigma v\rangle_{\textrm{fo}} \simeq \langle\sigma v\rangle_0$.
Then the additional contribution to the relic density can be obtained from
Eq.~(\ref{eq:Oh2contributions}). Examples of results of such a reconstruction (red squares) are shown in the left panel of Fig.~\ref{fig:Oh2nonth} for a $1$ TeV DM with the annihilation cross section equal to $\langle\sigma v\rangle_0 = 2\times 10^{-25}\,\textrm{cm}^3/\textrm{s}$. We assume that the DM particles annihilate either purely into $b\bar{b}$, $\tau^+\tau^-$, or into $b\bar{b}$ final state with $1\%$ admixture of $\gamma\gamma$. The mass reconstruction for the pure hadronic $b\bar{b}$ final state is quite poor, because the corresponding annihilation spectrum is often degenerate with the spectrum obtained for other pure or mixed final states for different DM masses. However, this can be drastically improved if either small admixture of the monochromatic line is present in the spectrum (from $\gamma\gamma$ or $Z\gamma$ final state; magenta dots) or the final state is not purely hadronic but also contains a leptonic contribution as can be seen for the case with $\tau^+\tau^-$ final state (blue triangles). On the other hand, it is important to note that the quality of the reconstruction of $\Omega_\chi^{\textrm{non-th}} h^2$ is rather sensitive to the derived values of $\langle\sigma v\rangle$ and therefore even a poor reconstruction of the mass parameter can lead to constraints on the non-thermal contribution to the DM relic density. 

However, if the DM particles interact more weakly, the quality of reconstruction decreases drastically leading to practically no conclusive results about the freeze-out origin of DM for $\langle\sigma v\rangle_0 = 1\times 10^{-25}$ (green squares) and $5\times 10^{-26}\,\textrm{cm}^3/\textrm{s}$ (grey squares) at a model-independent level. We will later see that this can be improved, \textsl{e.g.}, if a correlation between $\langle\sigma v\rangle$ and $\sigma_p^{\textrm{SI}}$ is derived for a specific model of the DM interactions.

Nonzero
values of $\alpha_p$ translate to $\langle\sigma v\rangle_{\textrm{fo}} > \langle\sigma v\rangle_{0}$, which suppresses the abundance
of WIMP DM at freeze-out and therefore enhances $\Omega^{\textrm{non-th}}_\chi h^2$. This is illustrated in the right panel of Fig.~\ref{fig:Oh2nonth} for two values of $\alpha_p/\alpha_s = 20$ and $100$ that lead to a non-negligible or even dominant $p$-wave contribution to the annihilation cross section  in Eq.~(\ref{eq:sigmavWIMP}) around the freeze-out temperature $T_{\textrm{fo}}\sim m_\chi/20$.

\begin{figure}[!ht]
\begin{center}
\includegraphics*[width=\textwidth]{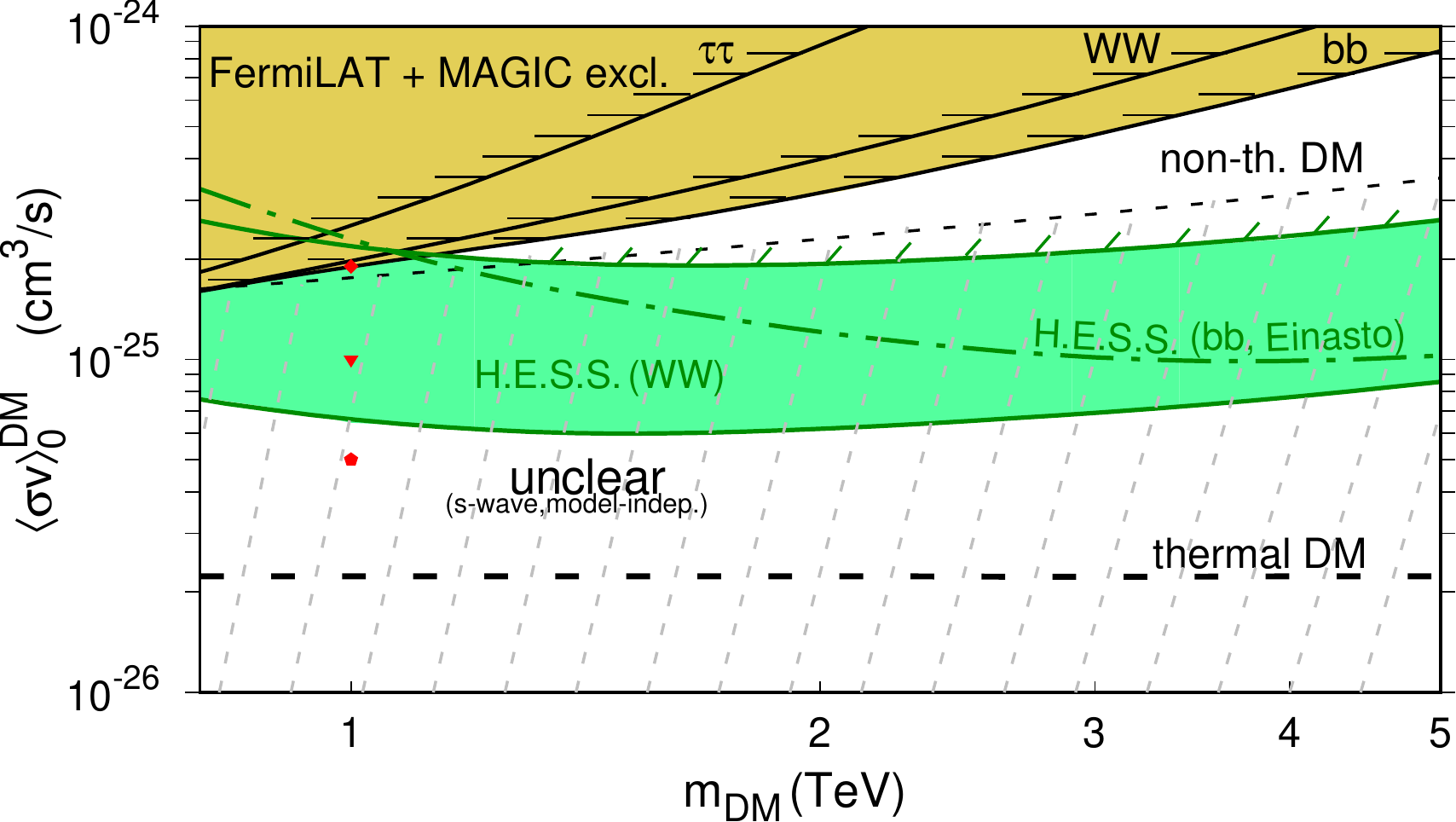}
\end{center}
\caption{Regions of the parameters of some underlying model (benchmark points) for which a non-zero value of the non-thermal contribution to the DM relic density, $\Omega_\chi^{\textrm{non-th}}h^2$ can be inferred. The region marked `unclear' corresponds to the benchmark points for which the 95\% CL range of the reconstructed values 
$\langle\sigma v\rangle_\mathrm{fo}$
contains the value of the annihilation cross section corresponding to the thermal WIMP DM relic density. The upper shaded region
represents the exclusion limits
of the FermiLAT and MAGIC collaborations depending on the final state of the DM annihilation, as discussed in \cite{Ahnen:2016qkx,Fermi-LAT:2016uux}.
The green lines represent the exclusion limits obtained by the H.E.S.S.\ collaboration, depending on the final state of the DM annihilation, as discussed in \cite{Abdallah:2016ygi}.
The spread of the line corresponding to the $WW$ final state reflects the uncertainty in DM halo profiles. The three benchmark points corresponding to $m_\chi=1\,\mathrm{TeV}$ and, respectively, to $\langle \sigma v \rangle_0=(2,\,1\,\,\textrm{and}\,\,0.5)\times10^{-25}\,\mathrm{cm}^3/\mathrm{s}$, are also shown.
}
\label{fig:simple}
\end{figure}

\begin{figure}[!ht]
\begin{center}
\includegraphics*[width=0.49\textwidth]{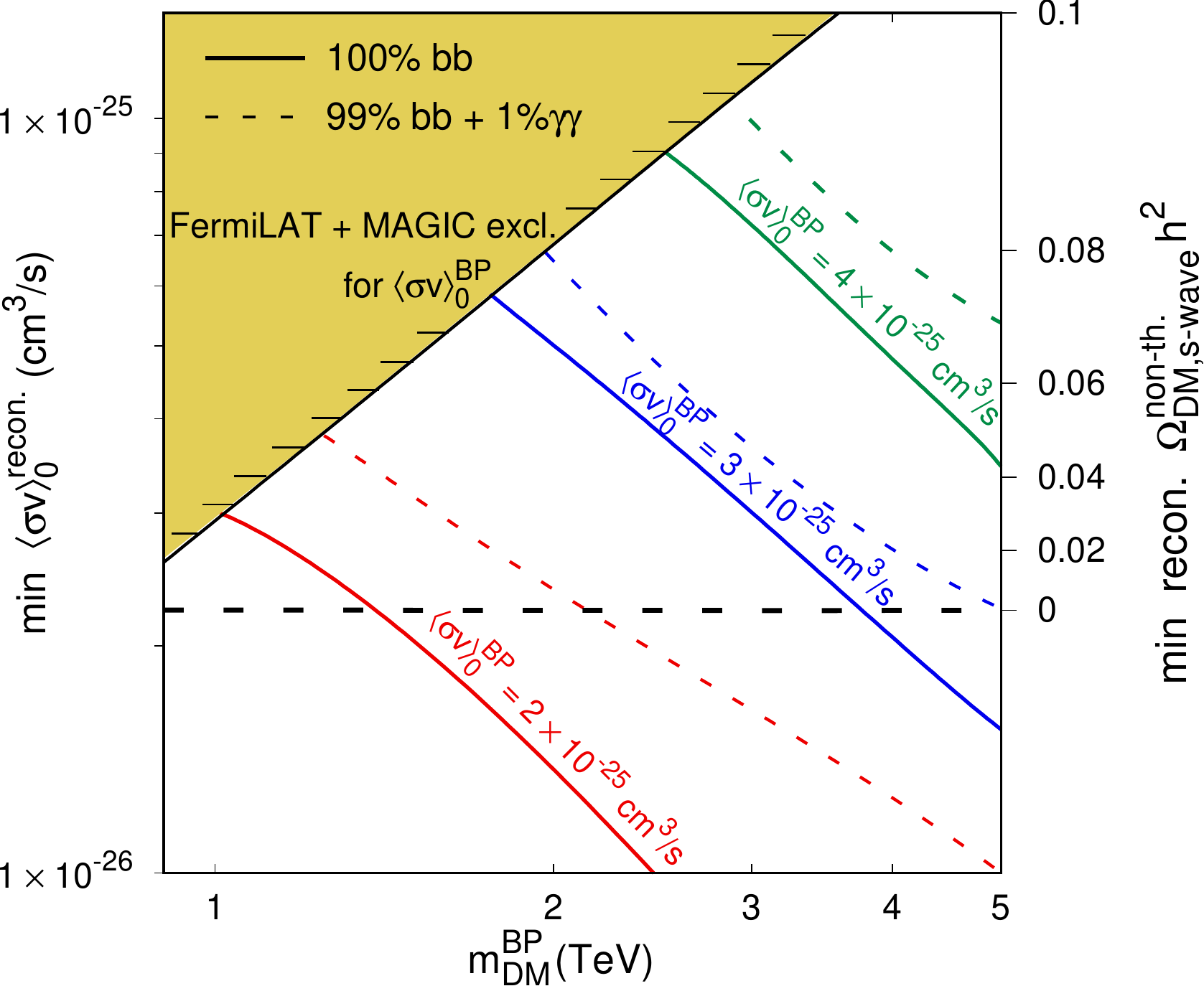}
\hfill
\includegraphics*[width=0.49\textwidth]{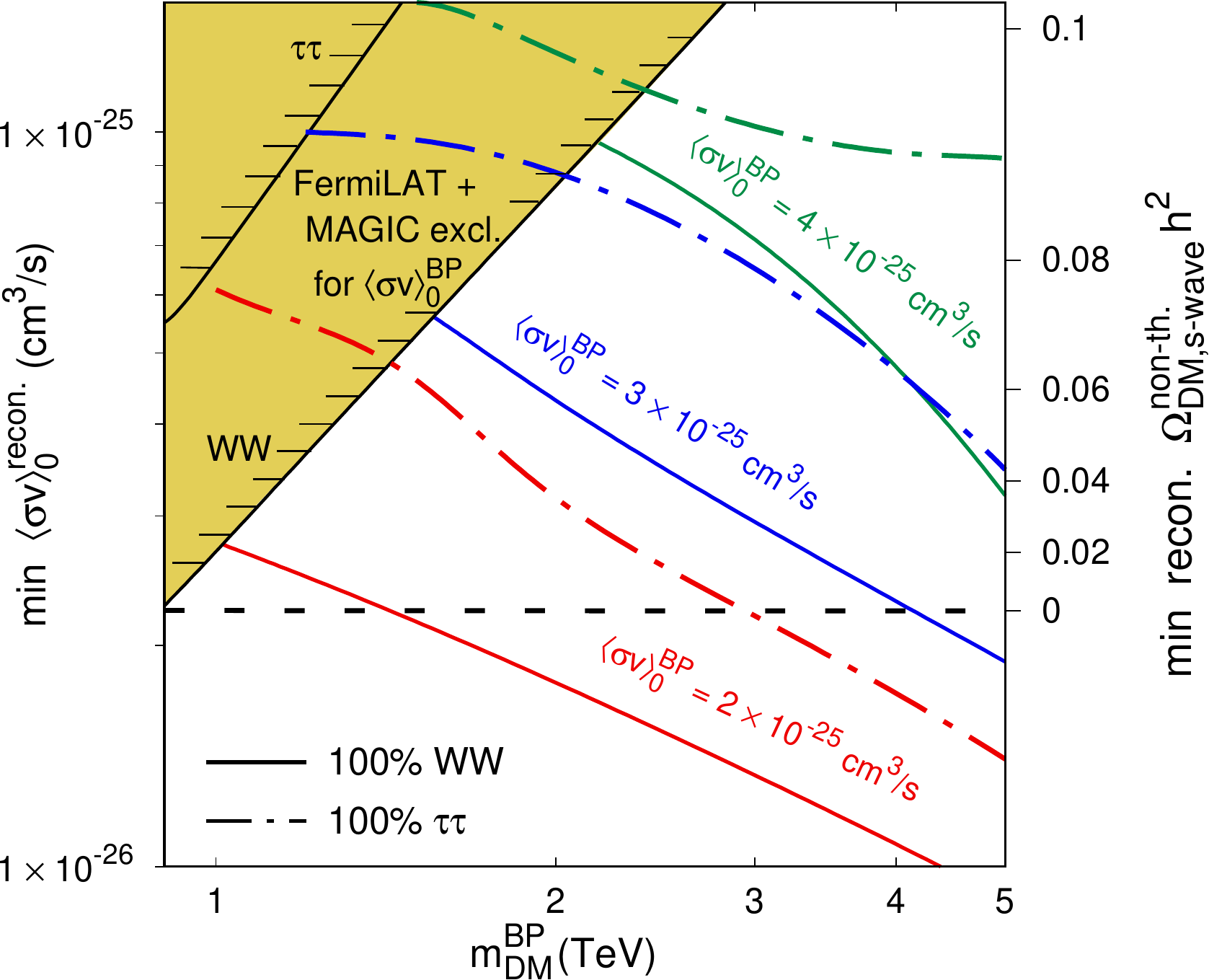}

\end{center}
\caption{Lower bound on the reconstructed value of the $\langle\sigma v\rangle_0^{\textrm{recon.}}$ as a function of the DM mass for the benchmark point, $m_\chi^{\textrm{BP}}$, for which the signal mock data set was generated for the reconstruction. A set of benchmark points that are considered corresponds to several values of the annihilation cross section, $\langle\sigma v\rangle_0^{\textrm{BP}}$, denoted in the plots. In the left panel the results obtained for pure $b\bar{b}$ final state (solid lines) and for $b\bar{b}$ final state with $1\%$ admixture of $\gamma\gamma$ channel (dashed lines) are shown. In the right panel we present the results for pure $W^+W^-$ (solid lines) and pure $\tau^+\tau^-$ (dash-dotted lines) final states. The bound on $\langle\sigma v\rangle_0$ is also translated into a limit on the non-thermal contribution to the DM relic density, $\Omega_{\chi,s-\textrm{wave}}^{\textrm{non-th}}$, assuming that the annihilation cross section around freeze-out is dominated by the $s$-wave contribution which is shown in the right side of each plot.}
\label{fig:Oh2sigmavbounds}
\end{figure}

These limits can be presented in a way that can be useful for future studies on the 
additional, non-thermal, DM production in the context of
specific models. 
In~Fig.~\ref{fig:simple}, we identify the values of $\langle\sigma v \rangle_0$ and $m_\chi$ of some underlying model (benchmark point) for which one
can infer that a non-thermal component to the WIMP DM density is necessary. The region marked `unclear' corresponds to the benchmark points for which the 95\% CL range of the reconstructed values 
$\langle\sigma v\rangle_\mathrm{fo}$
contains the value of the annihilation cross section 
consistent with the total WIMP DM density originating from freeze-out. 
We note that in a purely model-independent approach it is practically impossible to confirm the presence of non-thermal DM component, as large values
 of $\langle\sigma v\rangle_0$ that would allow ruling out the possibility of freeze-out-only DM are in conflict with ID searches except for tiny corners of the parameter space. However, it is important to remember that the published limits on $\langle\sigma v\rangle_0$, \textsl{e.g.}, those by the H.E.S.S. collaboration~\cite{Abdallah:2016ygi}, depend on the assumed DM profile which we allow to vary in our reconstruction (see Appendix~\ref{sec:methodology}).

With a specific choice of the annihilation final state, we can discuss the results in a finer way.
In Fig.~\ref{fig:Oh2sigmavbounds} we show the minimum value of the reconstructed $\langle\sigma v \rangle_{0,\textrm{min}}$ in the 95\%CL region as a function of the benchmark point WIMP DM mass, $m_\chi^{\textrm{BP}}$, for several choices of the annihilation cross section for the benchmark point, $\langle\sigma v \rangle_0^{\textrm{BP}}$. The limits are derived assuming that DM annihilates either purely into one of the final states that we consider or there is small $1\%$ admixture of the branching ratio into $\gamma\gamma$ final state that leads to a monochromatic contribution to the spectrum.
In the simplest case in which the DM annihilation in the early Universe is dominated by the $s$-wave contribution and coannihilations play negligible role, the minimum value $\langle\sigma v\rangle_{0,\textrm{min}}$ obtained in the reconstruction can be directly translated into a lower bound on $\langle\sigma v\rangle_{\textrm{f.o}}$ and therefore into limits on the contributions to the DM relic density can be derived. This is shown on the scales on the right-hand sides of the plots in Fig.~\ref{fig:Oh2sigmavbounds}, in which the minimal value of $\Omega_\chi^{\textrm{non-th}}h^2$ required to obtain the correct DM relic density is shown.
 These values can also be viewed as  conservative limits on $\Omega_\chi^{\textrm{non-th}}h^2$ since both non-negligible $p$-wave contribution, as well as additional coannihilations typically tend to increase $\langle\sigma v\rangle_{\textrm{fo}}$ above the current value $\langle\sigma v\rangle_0$.
 The annihilation cross section needs to be at least about an order of magnitude above the freeze-out value in order to allow the reconstruction of $\Omega_\chi^{\textrm{non-th}}h^2$, while for lower values one would not be able to determine if the additional contribution to the DM relic density is needed. However, as we will see in the next section, this can be improved within a framework of a specific model.

As can be seen in the left panel of Fig.~\ref{fig:Oh2sigmavbounds}, the quality of reconstruction of $\Omega_\chi^{\textrm{non-th}} h^2$ for $b\bar{b}$ final state can be improved if a small admixture of a monochromatic $\gamma\gamma$ line is present in the DM annihilation spectrum. Such an improvement is less pronounced for both $W^+W^-$ and $\tau^+\tau^-$ final states shown in the left panel of Fig.~\ref{fig:Oh2sigmavbounds}. In particular, in the case of $W^+W^-$ this is so because of the characteristic monochromatic-like spectral feature that is already present in the annihilation spectra. It comes from $W^\pm\rightarrow W^\pm \gamma$ splitting with the emission of soft $W^\pm$ and greatly increases the accuracy of DM mass determination.
For this reason we do not show the lines with small $\gamma\gamma$ admixture in the right panel of Fig.~\ref{fig:Oh2sigmavbounds}. In the plots we also show FermiLAT and MAGIC exclusion lines based on \cite{Ahnen:2016qkx}. It is important to note that these lines in Fig.~\ref{fig:Oh2sigmavbounds} cannot be directly compared to the usual exclusion lines 
shown in~\cite{Ahnen:2016qkx}, because in our case
they correspond to the benchmark values of the annihilation cross section, $\langle\sigma v\rangle_0^{\textrm{BP}}$, which are larger than the minimum values of the reconstructed cross section, $\langle\sigma v\rangle_0^{\textrm{recon.}}$. It is, however, the latter quantity that is shown on the vertical axes of both plots in Fig.~\ref{fig:Oh2sigmavbounds}, so the limits that we show in Fig.~\ref{fig:Oh2sigmavbounds} should be then rather thought as the lower limits on the DM mass for a given value of $\langle\sigma v\rangle_0^{\textrm{BP}}$ indicated in the plots.

The results presented above can be also used in the presence of invisible annihilation channels. Effectively, this is equivalent to having smaller $\langle\sigma v\rangle_0$ which leads to a poorer reconstruction of DM properties.


\subsection{More complicated examples}

\subsubsection{Multi-component thermal DM}

Scenarios of multi-component thermal DM pose an additional difficulty for extracting DM properties from the DM ID data even if (\ref{eq:sigmavWIMP}) is satisfied 
for each component separately. This is due to the fact that the DM ID rate is proportional to $\rho_i^2\langle\sigma v\rangle_{0,i}$, where $i$ runs over all DM 
components and $\rho_i$ is the local density of the $i$-th component, which in turn is proportional to the averaged annihilation cross-section for this component at freeze-out,
see Eqs~\ref{eq:DMflux} and \ref{eq:Jfac}. 
It is therefore possible that these cross-sections are larger than in the single-component (purely) freeze-out case and can be, in principle, detectable in the forthcoming DM ID experiments.

The prospects of determining DM properties are, however, significantly worse for these models compared to the case of single-component DM with both freeze-out 
and non-thermal contribution, because the DM ID rates scale as $(\rho_i/\rho_0)^2$, where $\rho_0$ is the present-day local DM density. We illustrate this
 in the left panel of Fig.~\ref{fig:2compDMSE}, where we show the reconstructed value of the annihilation cross section multiplied by the squared ratio of the local DM density of each DM component, $\rho_i$, with respect to the total local DM density, $\rho_0$. The benchmark point is the $b\bar{b}$, $1~\textrm{TeV}$ scenario discussed 
 in Section \ref{ssec:gen_sim}, with
$\langle\sigma v\rangle_0 = 2\times 10^{-25}~\textrm{cm}^3/\textrm{s}$.
 Light blue squares correspond to a scenario in which only one DM component leads to a $\gamma$-ray signal in DM ID. 
 We assume that this component has the same mass and cross-section as the benchmark model, but now only a thermal relic density
, \textsl{i.e.}, $\rho_i/\rho_0 = \Omega_{\chi,i}^{\textrm{fo}}h^2/0.12$.
 The green squares represent the reconstruction of the model in which there is one more DM component with thermal relic density and $m_\chi=500~\textrm{GeV}$, while $\langle\sigma v\rangle_{\textrm{fo}}$ ($s$-wave) is chosen such that the total DM relic density of both thermal DM components saturates the Planck data. In this case the observed $\gamma$-ray signal is a joint signal from both these DM particles. 
 In both examples, the quality of reconstruction is much worse than for the benchmark model alone; it resembles that of the single-component DM with a much smaller annihilation cross-section. 
%

\subsubsection{Sommerfeld enhancement}

A significant modification of the DM annihilation cross-section occurs in the non-relativistic regime when DM couples to a light mediator. This effect, known as Sommerfeld
enhancement \cite{Hisano:2002fk,Hisano:2003ec,Hisano:2004ds,Cirelli:2007xd,ArkaniHamed:2008qn}, invalidates the simple relation (\ref{eq:sigmavWIMP}) and we will discuss it separately.
Here,
we approximate the Yukawa potential by the Hulth\'en potential which allows to calculate the enhancement factor analytically~\cite{Cassel:2009wt,Slatyer:2009vg} (see also~\cite{Feng:2010zp,ElHedri:2016onc} for further discussion). We follow~\cite{Blum:2016nrz} to regularize velocity near a zero energy bound state and to
describe the mediator sector with two parameters: 
the coupling constant between the DM particle and light mediator, $\alpha_\chi$, and the mediator mass, $m_\phi$. When performing reconstruction, we allow them to vary in ranges $0.001 \leq\alpha_\chi\leq 1$ and $10^{-4} \leq m_\phi/m_\chi\leq 1$, respectively.

\begin{figure}[t]
\begin{center}
\includegraphics*[width=0.49\textwidth]{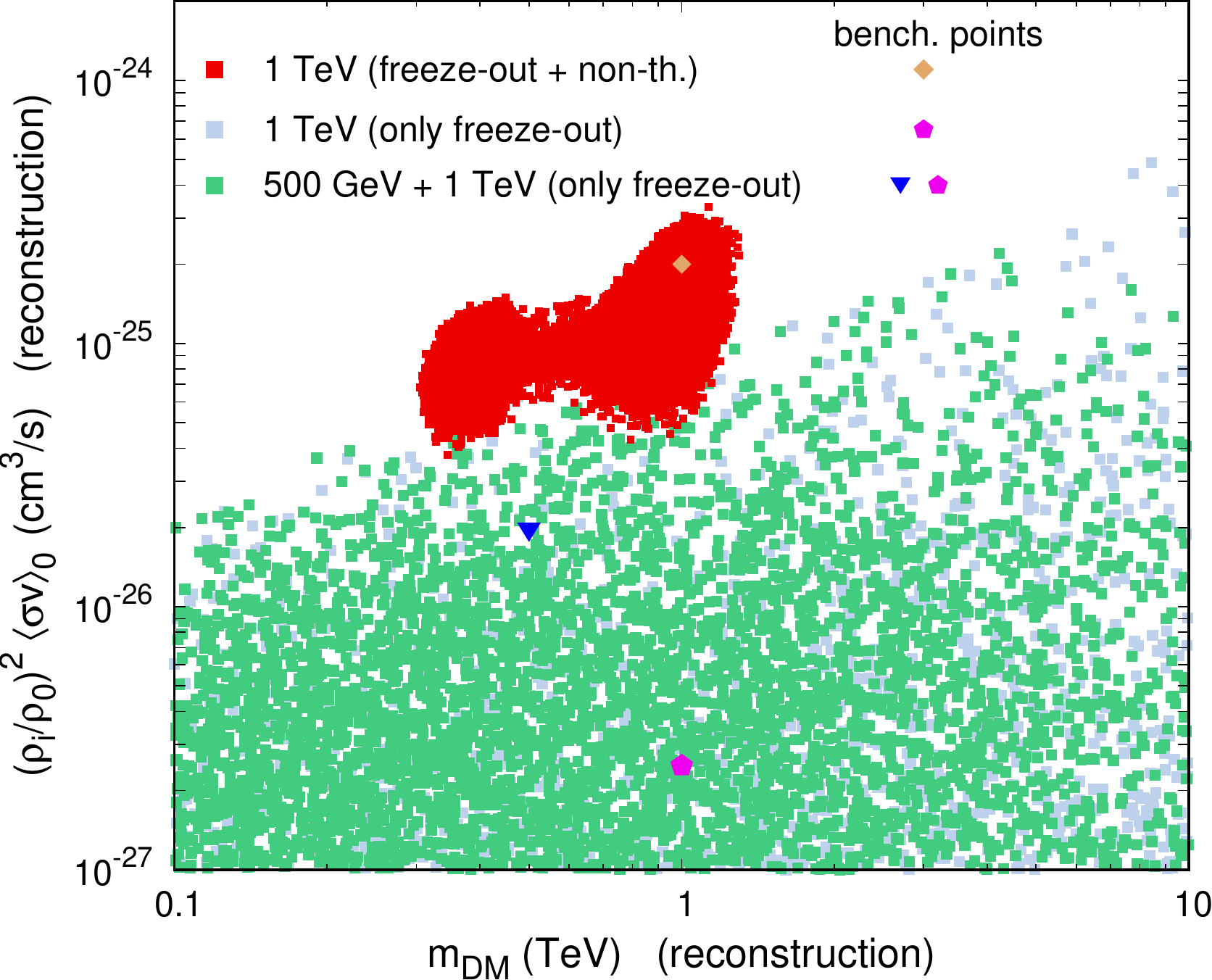}
\hfill
\includegraphics*[width=0.49\textwidth]{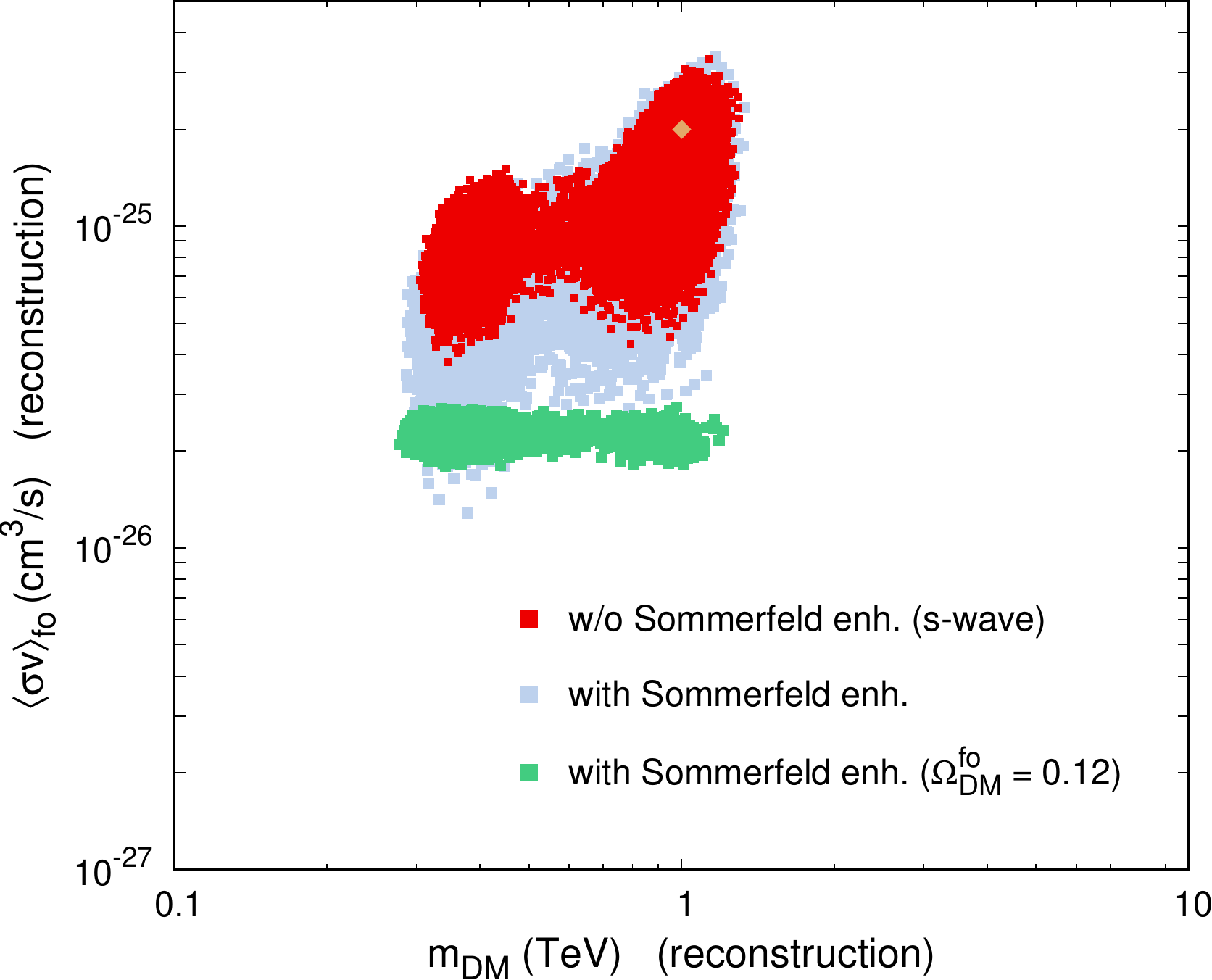}

\end{center}
\caption{Reconstruction for the benchmark scenario with $m_\chi=1~\textrm{TeV}$, $\langle\sigma v\rangle_0 = 2\times 10^{-25}\,\textrm{cm}^3/\textrm{s}$ and $b\bar{b}$ annihilation final state (light brown diamond). For a comparison on both panels we show (red squares) the reconstructed region for a single-component DM scenario without the Sommerfeld enhancement (SE) similarly to Fig.~\ref{fig:Oh2nonth}.
\textsl{Left panel}: Reconstruction for a multi-component DM scenario with the local DM density of each component denoted by $\rho_i$. We show the reconstructed value of the DM annihilation cross section, $\langle\sigma v\rangle_0$, multiplied by the squared ratio of $\rho_i$ with respect to the total local DM density, $\rho_0$, as a function of the reconstructed DM mass, $m_\chi$. The light blue squares correspond to a scenario in which $\gamma$-ray signal in DM indirect detection comes from only one DM component with thermal relic density which is characterized by the same $m_\chi$ and $\langle\sigma v\rangle_0$ as the aforementioned benchmark point. The green squares correspond to a scenario in which $\gamma$-ray signal additionally comes from the second type of DM particle assumed to have $m_\chi = 500~\textrm{GeV}$ and $\langle\sigma v\rangle_0 = 1\times 10^{-25}\,\textrm{cm}^3/\textrm{s}$ and thermal relic density. \textsl{Right panel}: Similar reconstruction of the annihilation cross section around freeze-out for a single-component DM scenario in the presence of the Sommerfeld enhancement of the annihilation cross section (see text for more details). The green (light blue) squares correspond to a scenario in which an additional constraint that the total DM relic density has a thermal origin is (not) imposed.}
\label{fig:2compDMSE}
\end{figure}

In the presence of the SE, the present value of the annihilation cross section, $\langle\sigma v\rangle_0$, can be larger than $\langle\sigma v\rangle_{\textrm{fo}}$, which means that the non-thermal contribution to DM density is smaller or not necessary at all. 
This is illustrated in the right panel of Fig.~\ref{fig:2compDMSE},
where the effects of SE for the benchmark model of Section \ref{ssec:gen_sim} are shown: we plot
the reconstructed value of the annihilation cross section around freeze-out 
for a fixed value of present-day annihilation cross-section.
Light green squares correspond to the assumption that the observed DM density has a fully thermal origin, while for light blue squares this requirement has not been imposed.
We can see that the effect can be large or small,
depending on the SE factor determined by $\alpha_\chi$ and $m_\phi/m_\chi$; in particular we can obtain the same $1~\textrm{TeV}$ BP for DM ID as discussed above, but with the $\langle\sigma v \rangle_{\textrm{fo}}$ around freeze-out at the thermal level for $m_\phi\simeq 10~\textrm{GeV}$ and $\alpha_\chi = 0.01$. 
We conclude that in the presence of SE it is impossible to infer whether a prospective observation of DM ID in forthcoming experiments signifies the need for non-thermal contribution to DM density. This is hardly surprising, as the MSSM wino is commonly known to be ruled out as a thermal DM candidate event with the current DM ID sensitivity \cite{Cohen:2013ama,Fan:2013faa,Hryczuk:2014hpa}.

Once the additional constraint on the thermal relic density is imposed, one expects the reconstruction of the annihilation cross section around freeze-out to be much better, as illustrated by light green points in the right panel of Fig.~\ref{fig:2compDMSE}. However, due to the freedom in obtaining large SE, the interplay between this strong constraint and the reconstruction based on DM ID signal is only via the reconstruction of the DM mass. Note that $m_\chi$ would remain practically unconstrained if only the DM relic density constraint was considered without DM ID signal taken into account.

%

\section{The effective field theory of DM: an example}
\label{sec:eft}

The quality of reconstruction of the annihilation cross section can be improved if, in addition to the signal coming from ID, DM is discovered in DD experiments thanks to possible correlations between scattering amplitudes in both types of searches \cite{Duda:2002hf}. Such correlations are, however, never ideal as typically $\langle\sigma v\rangle$ is governed by DM couplings to heavy fermions, \textsl{e.g.}, $b\bar{b}$ or $\tau^+\tau^-$, or gauge bosons, while DD rates are associated with DM couplings to light $u$ and $d$ quarks. 
Although limited, such correlations may improve the reconstruction within the framework of a given model,
because
fitting the same ID signal with different final states would require a shift in $\langle\sigma v\rangle_0$ and, correspondingly, in $\sigma_p^{\textrm{SI}}$, and the latter might be in tension with the simultaneous fitting of the DD signal. Such improved reconstruction of the annihilation final state can lead to better limits on the annihilation cross section and, subsequently, on the non-thermal contribution to the DM relic density.

We illustrate this effect within a framework of Effective Field Theory (EFT) approach to study DM couplings \cite{Beltran:2008xg,Cao:2009uw,Beltran:2010ww,Goodman:2010ku,Goodman:2010qn,Kumar:2013iva}. In particular, we assume vector-like couplings between DM and the SM particles
(see, \textsl{e.g.}, \cite{Goodman:2011jq,Frandsen:2012rk,Dreiner:2013vla,Buchmueller:2013dya,Abdullah:2014lla,Karwin:2016tsw,DEramo:2017zqw}). As we focus on the heavy DM particles, with $m_\chi \sim 1$ TeV, even with prospective ID and DD signals we typically remain beyond the reach of the LHC \cite{ATLAS:2012ky,Khachatryan:2014rra} (for EFT approach to DM DD see, \textsl{e.g.}, \cite{Fan:2010gt,Fitzpatrick:2012ix}).

We assume that the DM particles have vector-like effective couplings to the SM chiral eigenstates, $(c_f/\Lambda^2)\,(\chi\gamma^\mu\chi)(\widebar{f_{L/R}}\,\gamma_\mu\,f_{L/R})$, where $\Lambda$ corresponds to the cut-off scale of the UV completion of the model. The hierarchy between the coefficients $c_f$ (also assumed), is such that DM couples dominantly to 3rd generation fermions.
In the mass eigenstate basis this leads to the following Lagrangian that contains both vector and axial-vector couplings
\begin{eqnarray}
\mathcal{L}_{\textrm{eff}} & = & \frac{1}{\Lambda^2}\,\left(\bar{\chi}\gamma^\mu\chi\right)\,\Big\{\tilde{c}_{V,t}\,\bar{t}\gamma_\mu t + \tilde{c}_{V,b}\,\bar{b}\gamma_\mu b + \tilde{c}_{V,\tau}\,\bar{\tau}\gamma_\mu \tau + \tilde{c}_{V,\nu}\,\bar{\nu}_\tau\gamma_\mu \nu_\tau\nonumber\\
& & + \tilde{c}_{A,t}\,\bar{t}\gamma_\mu\gamma_5 t + \tilde{c}_{A,b}\,\bar{b}\gamma_\mu\gamma_5 b + \tilde{c}_{A,\tau}\,\bar{\tau}\gamma_\mu\gamma_5 \tau + \tilde{c}_{A,\nu}\,\bar{\nu}_\tau\gamma_\mu\gamma_5 \nu_\tau \Big\}.
\label{Eq:Lmass}
\end{eqnarray}
The Lagrangian in Eq.~(\ref{Eq:Lmass}) can be written in the chiral basis
\begin{equation}
\mathcal{L}_{\textrm{eff}} = \frac{1}{\Lambda^2}\,\left(\bar{\chi}\gamma^\mu\chi\right)\,\left\{c_{q,3}\,\widebar{q_L^{3}}\gamma_\mu q_L^3 + c_{u,3}\,\widebar{u_R^3}\gamma_\mu u_R^3 + c_{d,3}\widebar{d_R^3}\gamma_\mu d_R^3 + c_{l,3}\,\widebar{l_L^3}\gamma_\mu l_L^3 + c_{e,3}\,\widebar{e_R^3}\gamma_\mu e_R^3\right\}.
\label{Eq:Lelectroweak}
\end{equation}
with coefficients
\begin{equation}
c_{V/A,t} = \frac{\pm c_{q,3} + c_{u,3}}{2},\ \ \ c_{V/A,b} = \frac{\pm c_{q,3} + c_{d,3}}{2},\ \ \ c_{V/A,\tau} = \frac{\pm c_{l,3} + c_{e,3}}{2},\ \ \ c_{V/A,\nu} = \frac{\pm c_{l,3}}{2}. 
\label{Eq:couplings}
\end{equation}
In this effective model DM can then annihilate into one of the following final states or their mixture: $t\bar{t}$, $b\bar{b}$, $\tau^+\tau^-$ and $\nu_\tau\bar{\nu}_\tau$. The annihilation cross section would have both the vector and axial-vector contributions \cite{Srednicki:1988ce,Zheng:2010js}. Typically $p$-wave contributions to $\langle\sigma v\rangle$ are suppressed, but, for completeness, we take them into account (in addition to the dominant $s$-wave contributions) both when treating DM ID, as well as for the calculation of the DM relic density.

Although a direct coupling of the DM particles to $u$ and $d$ quarks is absent in Eq.~(\ref{Eq:Lmass}), the respective Wilson coefficients will be generated once the Renormalization Group (RG) evolution is taken into account from the renormalization scale $\Lambda$ down to the nuclear scale $\mu_N\sim 1$ GeV \cite{DEramo:2014nmf,DEramo:2016gos}. These Wilson coefficients for light $u$ and $d$ quarks determine the DD scattering cross section, $\sigma_p^{\textrm{SI}}$. The absence of direct couplings of DM to $u$ and $d$ quarks at the tree level allows us to obtain both $\langle\sigma v\rangle_0$ and $\sigma_p^{\textrm{SI}}$ that are large enough to be detected for the purpose of our reconstruction, but at the same time not yet excluded. 

\renewcommand{\arraystretch}{1.3}
\begin{table}[t]
   \centering\footnotesize
   \begin{tabular}{|c|c|} 
\hline
Parameters & Ranges\\
      \hline
      $m_\chi$ & $0.01-5$ TeV \\
      \hline
      $\Lambda$ & $1-10$ TeV \\
      \hline
	  $c_{q,3}$, $c_{u,3}$, $c_{d,3}$, $c_{l,3}$, $c_{e,3}$ & $0-1$\\
\hline
   \end{tabular}
   
   \caption{Parameters and their ranges to vary in the EFT model alongside with the nuisance parameters shown in Table~\ref{tab:nuiparams}.}
   \label{Tab:paramsEFT}
\end{table}

We vary coefficients $c_{q,3}$, $c_{u_3}$, $c_{d,3}$, $c_{l,3}$ and $c_{e,3}$, as well as the scale $\Lambda$ and the DM mass, $m_\chi$, as shown in Table~\ref{Tab:paramsEFT}.
 The reconstruction is performed for benchmark points characterized by the dominant annihilation channels into $b\bar{b}$ or $\tau^+\tau^-$. The DM mass, $m_\chi=1$ TeV, and the annihilation cross section, $\langle\sigma v\rangle_0$, are chosen at the same level as in the model-independent analysis. The $b\bar{b}$ ($\tau^+\tau^-$) benchmark point corresponding to the largest value of $\langle\sigma v\rangle_0 = 2\times 10^{-25}\,\textrm{cm}^3/\textrm{s}$ is characterized by $\sigma_p^{\textrm{SI}} = 2\times 10^{-46}\,\textrm{cm}^2$ ($3\times 10^{-46}\,\textrm{cm}^2$).
Our typical values of $\Lambda$ for benchmark points, as well as in the $95\%$ CL reconstructed regions, exceed $m_\chi$ by a factor of a few. The quality of the EFT approach in such a case has been studied, \textsl{e.g.}, in \cite{Buchmueller:2013dya} for axial-vector coupling: in such a scenario the EFT approach is either valid within some $20\%$ error or may lead to underestimated values of cross sections. In the latter case our effective model would correspond to the conservative lower limit on non-thermal contribution to the DM relic density which could be further improved if mediator is included in calculations.

\begin{figure}[t]
\begin{center}
\includegraphics*[width=0.49\textwidth]{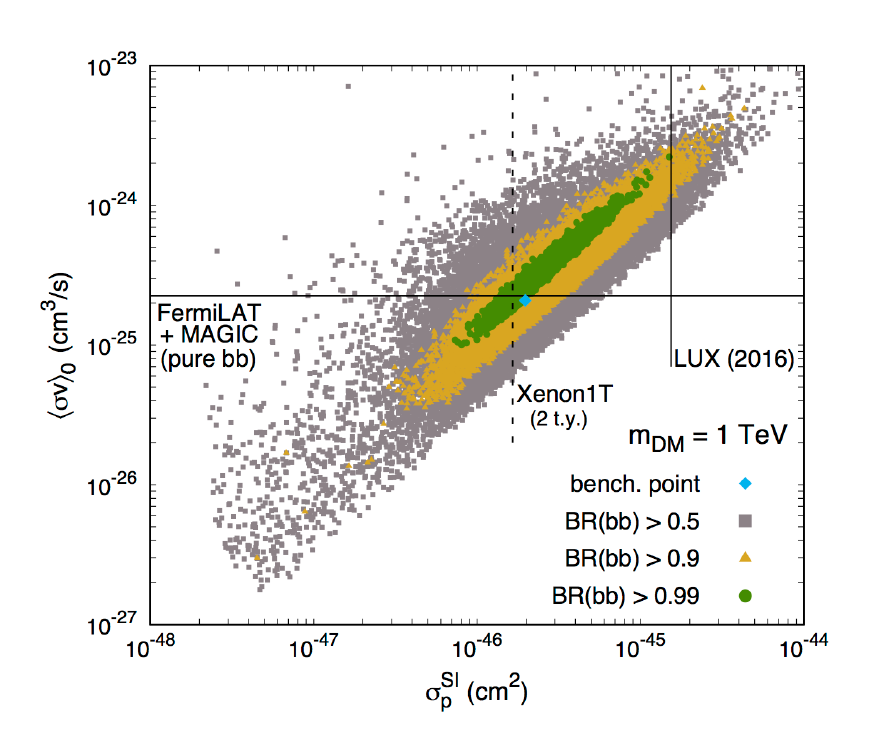}
\hfill
\includegraphics*[width=0.49\textwidth]{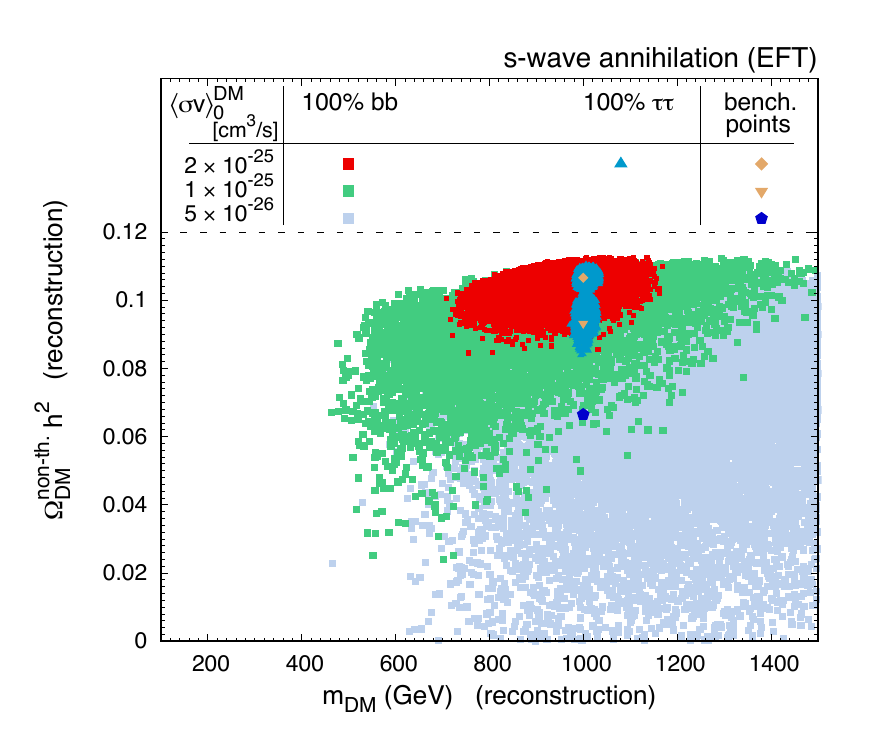}
\end{center}
\caption{\textsl{Left panel}: The correlation between spin-independent scattering cross section relevant for DD, $\sigma_p^{\textrm{SI}}$ and DM annihilation cross section, $\langle\sigma v\rangle_0$ for the EFT model. The points in the plot correspond to the DM mass, $m_\chi = 1$ TeV and increasing annihilation branching ratio into $b\bar{b}$ quarks. Solid horizontal line corresponds to the combined FermiLAT and MAGIC exclusion limit \cite{Ahnen:2016qkx} for pure $b\bar{b}$ final state. The current limit from LUX collaboration \cite{Akerib:2016vxi} is shown as a vertical solid line, while projected sensitivity obtained for 2 tonne-years exposure of the Xenon1T experiment \cite{Aprile:2015uzo} corresponds to the vertical dashed line. \textsl{Right panel}: Reconstruction of the non-thermal contribution to the DM relic density for the EFT model for the benchmark point annihilates dominantly into $b\bar{b}$ (squares) or $\tau^+\tau^-$ (blue triangles). The position of the benchmark points correspond to $m_\chi = 1$ TeV and three values of $\langle\sigma v\rangle_0 = (2,\,1\,\,\mathrm{and}\,\,0.5)\times 10^{-25}\,\textrm{cm}^3/\textrm{s}$. The $b\bar{b}$ ($\tau^+\tau^-$) benchmark point corresponding to the largest value of $\langle\sigma v\rangle_0$ is characterized by $\sigma_p^{\textrm{SI}} = 2\times 10^{-46}\,\textrm{cm}^2$ ($2\times 10^{-46}\,\textrm{cm}^2$). The horizontal black dashed line corresponds to the total DM relic density. 
}
\label{fig:Oh2nonthEFT}
\end{figure}

In the left panel of Fig.~\ref{fig:Oh2nonthEFT} we show the correlation between $\sigma_p^{\textrm{SI}}$ and $\langle\sigma v\rangle_0$ for points in the considered EFT model with $m_\chi = 1$ TeV and varying annihilation branching ratio to $b\bar{b}$ final state. In particular, the points with $\textrm{BR}(b\bar{b})>0.99$ (green circles) shown strong correlation. This is not surprising, since for the fixed mass and annihilation cross section, both $\sigma_p^{\textrm{SI}}$ and $\langle\sigma v\rangle_0$ effectively depend on a single quantity, \textsl{i.e.}, the cut-off scale $\Lambda$. On the other hand, once one allows more efficient annihilation into different final states, the correlation becomes weaker. 

We can now see how an interplay between DD and ID experiments can help to improve reconstruction of the DM properties. In the large mass regime, the DD experiments have no sensitivity to DM mass, as discussed in Appendix~\ref{Sec:DD}, 
but
if the DM signal is observed in the ID experiment
one can obtain relatively accurate values of $m_\chi$ and the dominant annihilation branching ratio. This can lead to a stronger correlation in expected signal between ID and DD experiments, which, in turn, results in an improved reconstruction of both $\sigma_p^{\textrm{SI}}$ and $\langle\sigma v\rangle_0$.

Finally, one can translate the improved limit on $\langle\sigma v\rangle_0$ into the limit on the non-thermal contribution to the DM relic density. We show such an improved reconstruction of $\Omega^{\textrm{non-th.}}_\chi h^2$ in the right panel of Fig.~\ref{fig:Oh2nonthEFT} (compare with the left panel of Fig.~\ref{fig:Oh2nonth} for which we used the same $m_\chi$ and $\langle\sigma v\rangle_0$ for the benchmark points).
Compared to the analysis presented in Section \ref{sec:gen}, we note
an improved mass reconstruction in the ID experiments with respect to the similar analysis discussed in the model-independent section, which comes partly from a smaller set of final states in the considered EFT model (specifically, the lack of $W^+W^-$ and $hh$ final states). On the other hand the improvement in the reconstruction of $\Omega_\chi^{\textrm{non-th.}}h^2$ is to a large extent the effect of the correlation between the ID and DD experiments.


\section{Neutralino DM\label{sec:neutralinoDM}}
\label{sec:neutralino}

In the preceding Sections, we have shown what we can learn about the generation of WIMP DM  
without invoking specific assumptions about the
underlying microscopic model of DM. We then discussed how such a reconstruction can be improved by adding simple assumptions about the correlations between the DD and ID rates. We illustrated the latter feature within a framework of a simple EFT model. In both cases we focused solely on DM itself, \textsl{i.e.}, we ignored any possible additional contribution to $\langle\sigma v\rangle_{\textrm{fo}}$ from coannihilations that could become important if the DM particles are mass-degenerate with some other species in the thermal plasma in the early Universe. 

By taking into account coannihlations one allows in general at least one more degree of freedom associated with the mass difference between $\chi$ and the heavier species. This, in principle, could make it much more difficult to draw any conclusions about the contributions to the DM relic density from reconstruction of the current value of the annihilation cross section, $\langle\sigma v\rangle_0$, unless some other non-trivial relations between all these quantities are present in the model. We will illustrate this point by performing a study of a prototypical example of a WIMP DM particle, \textsl{i.e.}, the lightest neutralino, $\chi\equiv\chi_1^0$, which appears in the context of supersymmetric theories. 
 
\subsection{Reconstruction of non-thermal component of DM with wino and \mbox{higgsino} DM}
 
In general, the lightest neutralino is a linear combination of the superpartners of  the gauge bosons and the Higgs bosons. In our study we focus on scenarios in which the neutralino is higgsino or wino dominated. It is known that for these type of DM particles freeze-out can provide a correct amount of 
DM density
for about $m_\chi\sim 1$ TeV and $2-3$ TeV, respectively, while for lower DM mass one obtains too low $\Omega_\chi^{\textrm{fo}}h^2$,
so a discovery of such a WIMP DM particle at a lower mass scale would signify the necessity of the non-thermal contribution to $\Omega_\chi h^2$ (or another DM species).

We also note that in both scenarios coannihilations play an important role in establishing the DM relic density in the early Universe. In particular, the difference in mass between $\chi$ and the lighter chargino, as well as second to the lightest neutralino in the case of higgsino DM, is typically bound to no more than several GeV. For such a small mass difference, the DM freeze-out density is not very sensitive to the precise determination of the mass degeneracy, but it depends on the model parameters that also determine the composition of the lightest neutralino which in turns settles the actual values of both $\langle\sigma v\rangle$ and $\sigma_p^{\textrm{SI}}$. Such a set of non-trivial correlations between DD and ID rates, as well as the DM relic density allows for successful reconstruction of the non-thermal component of the DM relic density even in the presence of coannihilations during thermal freeze-out.


We present our results for neutralino DM for two benchmark points that correspond to higgsino DM with $m_\chi \simeq 400$ GeV and (predominantly) wino DM with $m_\chi\simeq 1.1$ TeV. The cross sections for direct and indirect detection for both benchmark points have been chosen to lie just below
the current upper bounds obtained by the LUX Collaboration \cite{Akerib:2016vxi}, FermiLAT + MAGIC \cite{Ahnen:2016qkx,Fermi-LAT:2016uux} and H.E.S.S\cite{Abdallah:2016ygi}, respectively. It is important to note that these bounds should not be treated as a strict limits since they correspond to the standard choice of the astrophysical parameters while we allow them to vary in the scan (see Table~\ref{tab:nuiparams}).

\begin{figure}[t]
\begin{center}
\includegraphics*[width=0.49\textwidth]{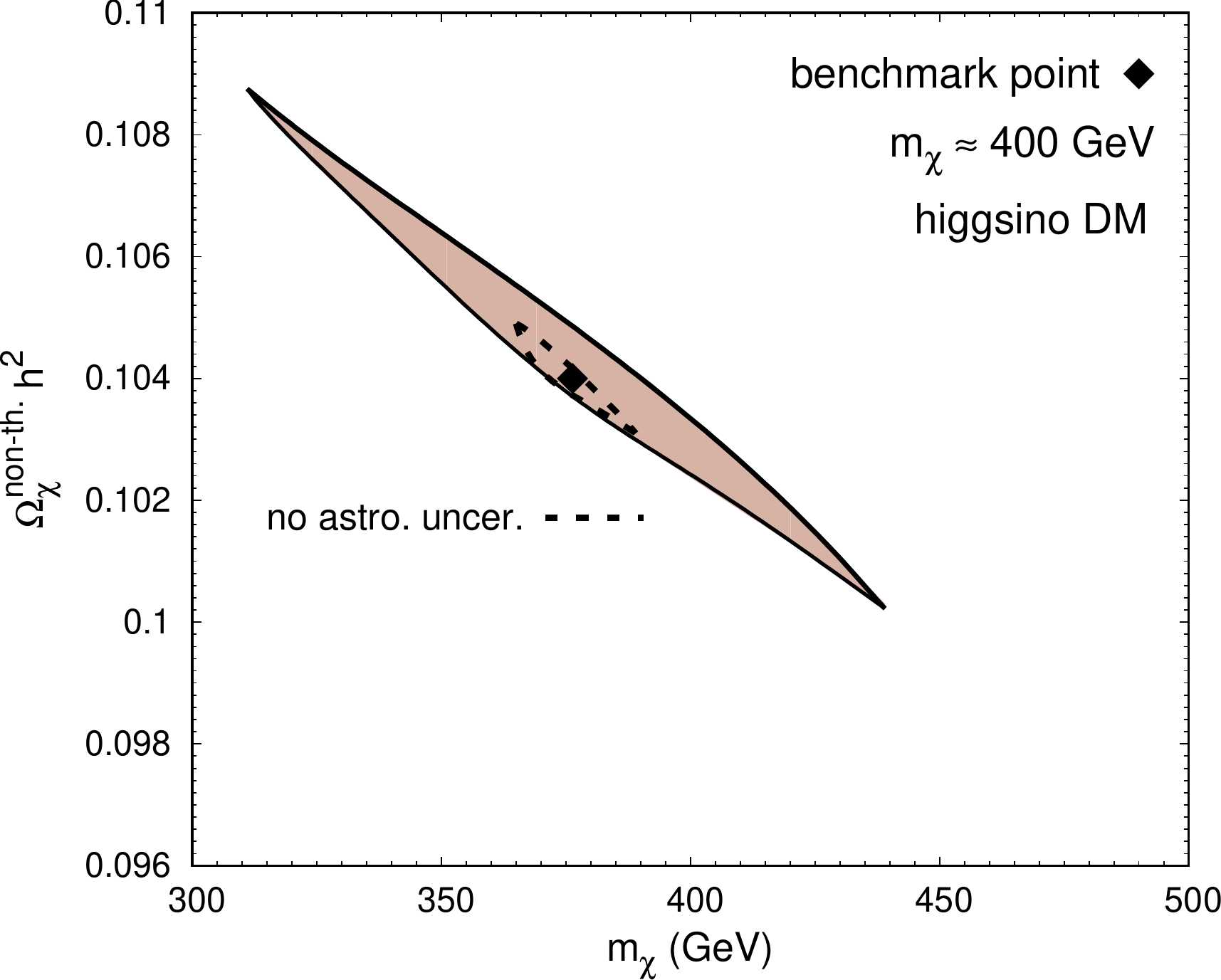}
\hfill
\includegraphics*[width=0.49\textwidth]{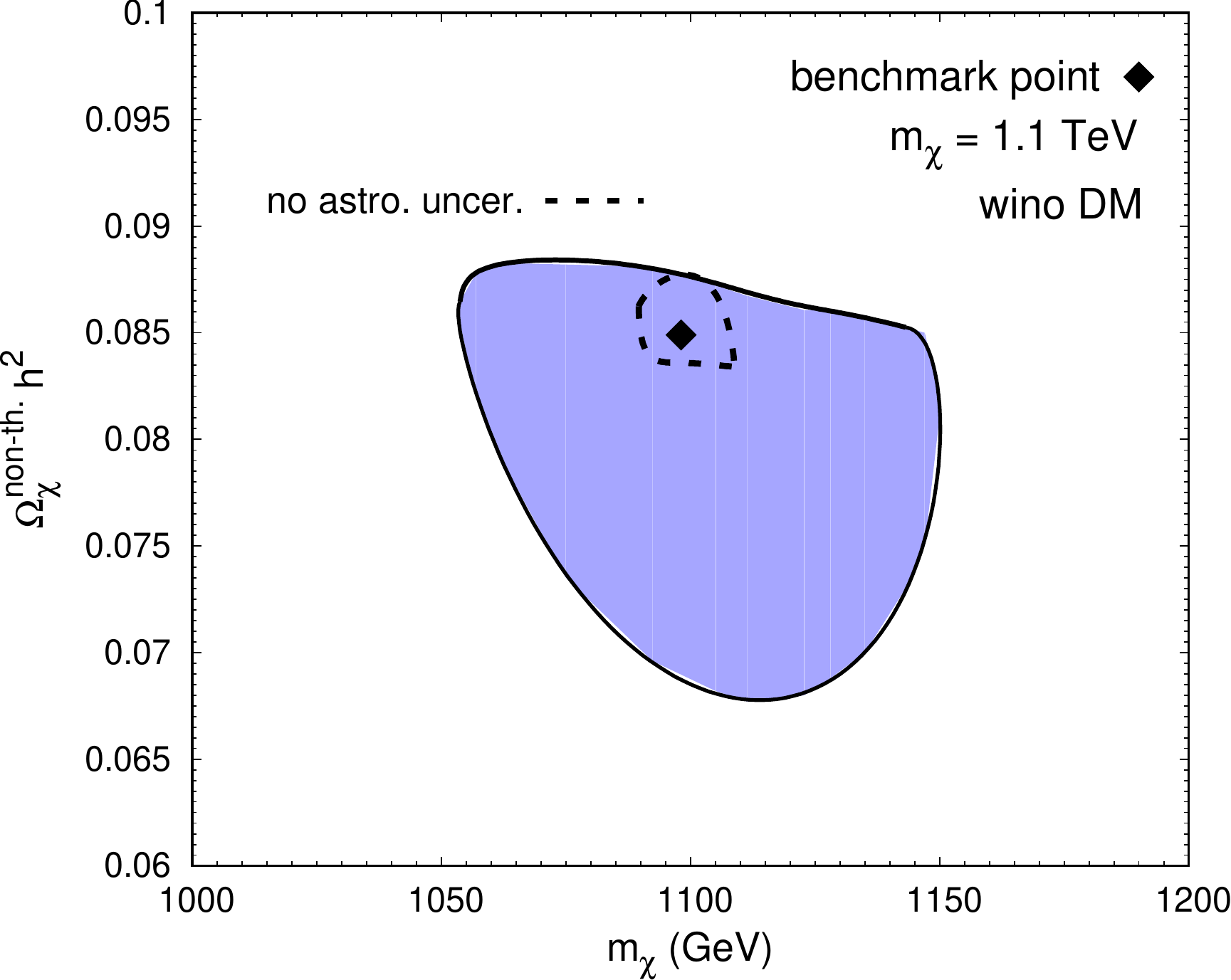}

\end{center}
\caption{The quality of reconstruction ($95\%$ CL regions) of the non-thermal contribution to the DM relic density, $\Omega_\chi^{\textrm{non-th}}h^2$, as a function of the reconstructed mass of DM, $m_\chi$, for the benchmark points (marked by black diamond) corresponding to higgsino DM (left panel) with $m_\chi = 400$ GeV and wino DM (right panel) with $m_\chi = 1.1$ TeV. Dashed lines enclose the $95\%$ CL regions obtained when the astrophysical uncertainties are neglected.} \label{fig:Oh2neutralino}
\end{figure}


We perform a numerical analysis of a 10-parameter phenomenological MSSM; details of the numerical procedure and the applied constraints are 
described in the Appendix~\ref{App:B}.
The $95\%$ CL regions 
obtained after fitting to the particle physics constraints 
and to the DM signal mock data set are shown in Fig.~\ref{fig:Oh2neutralino} for both of our benchmark points. We present the results in the $(m_\chi,\Omega_\chi^{\textrm{non-th}}h^2)$ plane where $\Omega_\chi^{\textrm{non-th}}h^2 = 0.12 - \Omega_\chi^{\textrm{fo}}h^2$.  The quality of reconstruction allows to constrain the necessary non-thermal contribution to the neutralino DM relic density, which varies between $\Omega^{\textrm{non-th}}_\chi h^2 = 0.1$ and $0.11$ for the higgsino DM case, and between $0.07$ and $0.09$ for the wino DM scenario. The upper limits of the reconstructed $95\%$ CL regions correspond to smaller freeze-out contribution,
which can be roughly translated to larger $\langle \sigma v \rangle_0$,
implying that the current constraints from DM ID 
become important.
The lower limits on $\Omega_\chi^{\textrm{non-th}}h^2$ are driven by the quality of reconstruction of the  DM ID 
signal,
which is mainly affected by astrophysical uncertainties: a change in the nuisance parameters described in Table~\ref{tab:nuiparams} can
make a model with smaller $\langle \sigma v \rangle_0$ mimic the DM ID signal.
For both our benchmark points, the $95\%$ CL regions would be significantly smaller if one neglected the astrophysical uncertainties.

The absolute quality of the DM mass reconstruction 
is similar for both benchmark points, irrespective of $m_\chi$.
This follows from a more pronounced monochromatic-like spectral feature coming from $W^+W^-$ final state for $m_\chi \gg m_W$. 
On the other hand, 
$\Omega_\chi^{\textrm{non-th}}h^2$ is better reconstructed for the higgsino DM benchmark point with $m_\chi^{\textrm{BP}} = 400$ GeV. It is mainly because of the interplay between the DM DD and ID which 
limits the impact of varying $\rho_\chi$. 
As can be seen from Eqs~(\ref{eq:DMflux}), (\ref{eq:Jfac}) and (\ref{eq:dRdE}), modifying $\rho_\chi$ in such a way that $\rho_\chi^2\langle\sigma v\rangle_0$ remains unchanged 
requires a change in $\sigma_p^{\textrm{SI}}$ such that $\rho_\chi\sigma_p^{\textrm{SI}}$ is roughly constant. For a given $m_\chi$ this can be achieved 
by adjusting the gaugino component of the higgsino. 
In principle, there are two adjustable gaugino mass parameters, \textsl{i.e.}, the bino mass, $M_1$, and the wino mass, $M_2$, which appears to give enough freedom to simultaneously fit ID and DD signals for modified $\rho_\chi$. However, increasing the bino component, 
would lead to a decrease of the annihilation cross section and an increase of $\sigma_p^{\textrm{SI}}$. 
Therefore, varying the bino component does not help in `hiding' the DM signal in both DD and ID; since for a varying wino component the DD and ID constraints are non-degenerate, the DM reconstruction becomes more accurate in the higgsino DM case.

\subsection{Origin on non-thermal DM and constraints on the reheating temperature}

In the previous section we studied the quality of reconstruction of the non-thermal contribution to the DM relic density of higgsino and wino DM without specifying its actual origin. 
Interestingly, within the MSSM there exists a mechanism for the production of such additional DM particles; it involves
extremely weakly interacting massive particles (EWIMPs), with interactions so weak that
EWIMPs decay to DM particles long after the freeze-out.
Theoretically motivated EWIMPs include the axino \cite{Nilles:1981py,Frere:1982sg,Tamvakis:1982mw,Covi:1999ty,Covi:2001nw} (see also \cite{Kim:2012bb} and for a review \cite{Baer:2014eja,Choi:2013lwa}) and the gravitino \cite{Deser:1977uq,Cremmer:1978iv,Cremmer:1978hn,Cremmer:1982en,Feng:2003xh,Bolz:2000fu} that appear in supersymmetric models.
%
%
Both EWIMPs have very low interaction rates, suppressed by a high cutoff scale, but nonetheless they
can be effectively produced in, \textsl{e.g.}, scatterings and/or decays of other particles that were in thermal equilibrium themselves \cite{Covi:2001nw,Bolz:2000fu,Brandenburg:2004du,Pradler:2006qh,Rychkov:2007uq,
Bae:2011jb,Choi:2011yf,Bae:2011iw}. 

EWIMPs interact too weakly to be detected in any of the current or foreseeable future experiments searching for particles beyond the SM. Typically, the limits on these particles come from cosmological considerations about the Big Bang Nucleosynthesis (BBN) \cite{Reno:1987qw,Kawasaki:1994sc,Feng:2003uy,Feng:2004zu}, Large Scale Structure (LSS) formation \cite{Jedamzik:2005sx} and Cosmic Microwave background (CMB) radiation \cite{Ellis:1984eq,Hu:1993gc}. However,  within a given model, one can look for collider signals associated with species heavier than EWIMP, see, \textsl{e.g.}, in \cite{Ambrosanio:1996jn,Buchmuller:2004rq,Covi:2004rb,Feng:2004mt,
Roszkowski:2004jd,
Hamaguchi:2004df,Feng:2004yi,Brandenburg:2005he,
Choi:2007rh,Feng:2010ij} (for more recent studies related to this topic see, \textsl{e.g.}, \cite{Aad:2014gfa,Maltoni:2015twa,Aad:2015zva,Khachatryan:2015vta,
Arvey:2015nra,Ibe:2016gir}). 
Alternatively, additional information can be obtained from DM ID signals in the scenarios of decaying gravitino or axino DM \cite{Bomark:2014yja,Kong:2014gea,Choi:2014tva,Liew:2014gia,Arcadi:2015ffa,
Gomez-Vargas:2016ocf}.
Altogether, these constraints on DM particles and their EWIMP companions translate to constraints on the thermal history of the Universe.

We will illustrate this issue for the neutralinos that come from late-time decays of heavier 
axinos or gravitinos after neutralino freeze-out.
In that case, eq.~(\ref{eq:Oh2contributions}) can be written as
\begin{equation}
\Omega_\chi h^2 
= \Omega_\chi^{\textrm{fo}}h^2 + \frac{m_{\chi}}{m_{\textrm{EWIMP}}}\,\Omega_{\textrm{EWIMP}}^{\textrm{TP}}h^2 = \Omega_\chi^{\textrm{fo}}h^2 + c\,m_\chi\, Y_{\textrm{EWIMP}},
\label{eq:Oh2contributions2}
\end{equation}
where 
 $Y^{\textrm{TP}}_{\textrm{EWIMP}}$ denotes the yield of EWIMPs from so-called thermal production of EWIMPs,\footnote{Thermal production of EWIMPs should not be confused with freeze-out thermal production of WIMPs. In the latter case one assumes that WIMPs are in thermal equilibrium in the early Universe. This is, however, not true for EWIMPs for which one only assumes that they originate from decays and scatterings of other species that are in thermal equilibrium.}, $c = s_0/\rho_c = 2.741\times 10^8\,\textrm{GeV}^{-1}$, where $s_0 = 2889.2\,\textrm{cm}^{-3}$ is the present-day entropy density and $\rho_c = 1.0539\times 10^{-5}\,\textrm{GeV\,cm}^{-3}$ is the critical density, and we assume $m_{\textrm{EWIMP}}> m_\chi$. 
For the gravitino or the axino, their production from thermal plasma is governed by non-renormalizable operators and therefore the EWIMP yield is sensitive to the reheating temperature, $T_R$, which can be identified with the maximal temperature in the radiation dominated (RD) epoch in the evolution of the Universe.\footnote{For a recent discussion of corrections to the axino yield due to non-instantaneous reheating period see \cite{Roszkowski:2015psa}. In the case of the renormalizable operators the final yield of EWIMPs would be proportional to the annihilation cross section \cite{Giudice:2000ex}, $Y_{\textrm{EWIMP}}\sim \langle\sigma v\rangle$, as 
in any generic freeze-in mechanism \cite{Hall:2009bx}.}

\begin{figure}[!t]
\begin{center}
\includegraphics*[width=0.49\textwidth]{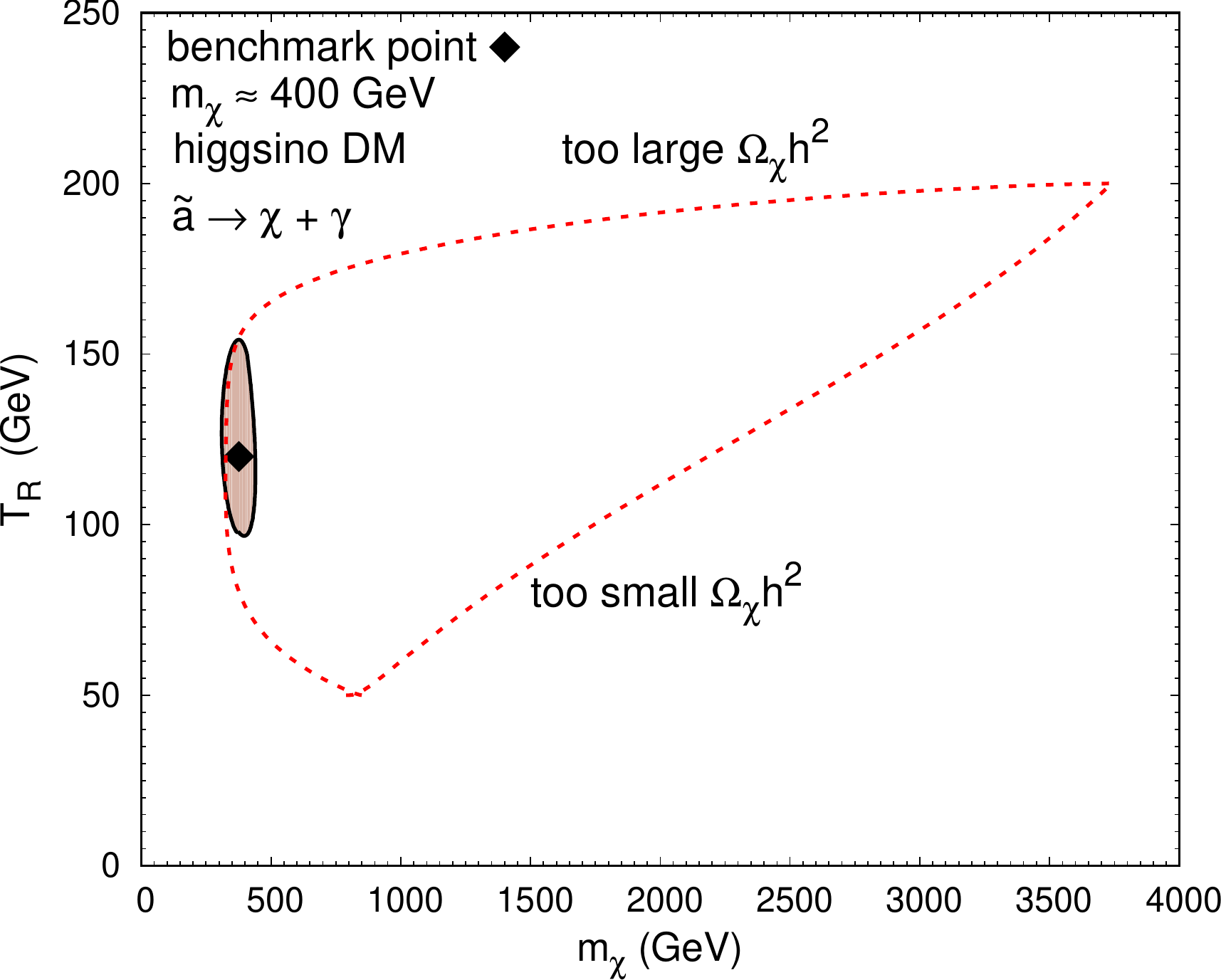}
\hfill
\includegraphics*[width=0.49\textwidth]{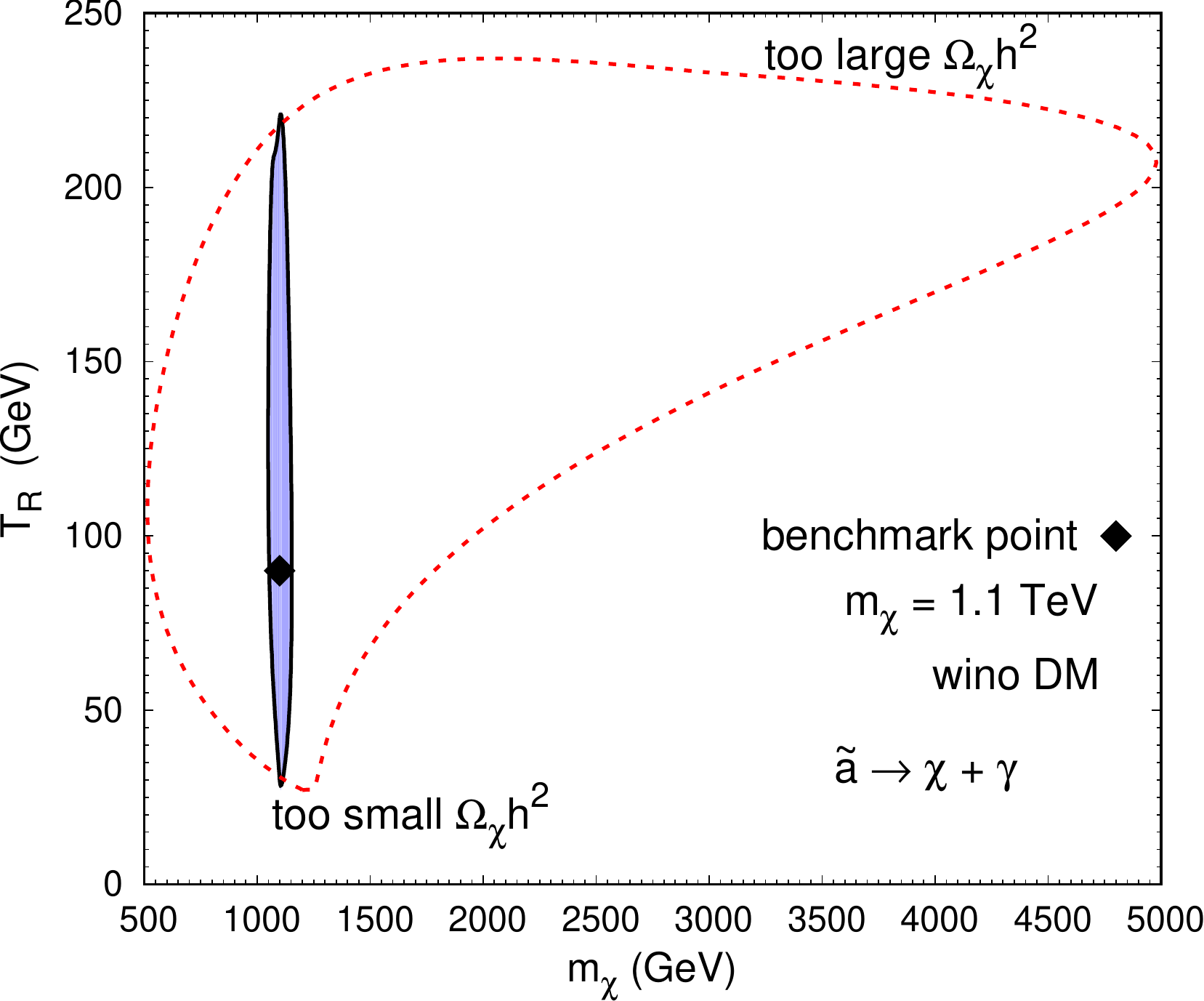}
\end{center}
\caption{The quality of reconstruction of the reheating temperature, $T_R$, as a function of the reconstructed mass of DM, $m_\chi$, for the same benchmark points as used in Fig.~\ref{fig:Oh2neutralino} for scenario with heavy axino decaying into higgsino (left panel) or wino (right panel) DM. 
Red dashed lines mark the boundary of the region consistent with particle physics constraints, \textsl{i.e.}, without the signal from a prospective DM
discovery taken into account.
} \label{fig:TRreconneutralino}
\end{figure}

\subsubsection{Axino EWIMP}

Axinos can be produced effectively if the reheating temperature of the Universe after inflation, $T_R$, is smaller than the temperature at which these EWIMPs establish thermal equilibrium or they could otherwise easily overclose the Universe. The amount of produced axino EWIMPs is typically proportional to $T_R$, unless the reheating temperature becomes small and the Boltzmann suppression of the number densities of the involved particles needs to be taken into account. 
For values of $T_R$ smaller than the freeze-out temperature one also needs to consider a modified evolution
of the Universe around freeze-out.
We take all this effects into account following \cite{Roszkowski:2015psa,Roszkowski:2014lga}. We consider the supersymmetrized version of Kim-Shifman-Vainstein-Zakharov type of axion models \cite{Kim:1979if,Shifman:1979if}.
We also assume that the axino is lighter than the gluino, $m_{\tilde{g}} > m_{\tilde{a}}>m_\chi$.
This is merely a technical assumption: otherwise, depending whether the gluinos decay into neutralino DM before or after freeze-out, this extra contribution
to DM is either erased or stays intact, respectively \cite{Choi:2008zq}. As a result, we obtain one of two simple cases, with the choice sensitive to the details of the MSSM spectrum.
With this assumption, the axino mass can be varied between $m_\chi$ and $m_{\tilde{g}}$ without affecting much the non-thermal contribution to the neutralino relic density which depends essentially only on $T_R$ for a given point in the parameter space of the MSSM. 
Therefore,
we can present previously derived bounds on $\Omega_\chi^{\textrm{non-th}}h^2$ as constraints on $T_R$ as shown in 
Fig.~\ref{fig:TRreconneutralino}. As expected the reheating temperature in this scenario is confined to low values and 
the order of magnitude of
its value can be determined 
for both our benchmark points. The uncertainty on $T_R$ is mainly driven by the lack of accuracy in determination of $\Omega_\chi^{\textrm{non-th}}h^2$ and $m_\chi$, but to some extent it also comes  from the fact that the production of axinos for such low $T_R$ is sensitive to the rest of the supersymmetric spectrum.

\subsubsection{Gravitino EWIMP}

\begin{figure}[!t]
\begin{center}
\includegraphics*[width=0.49\textwidth]{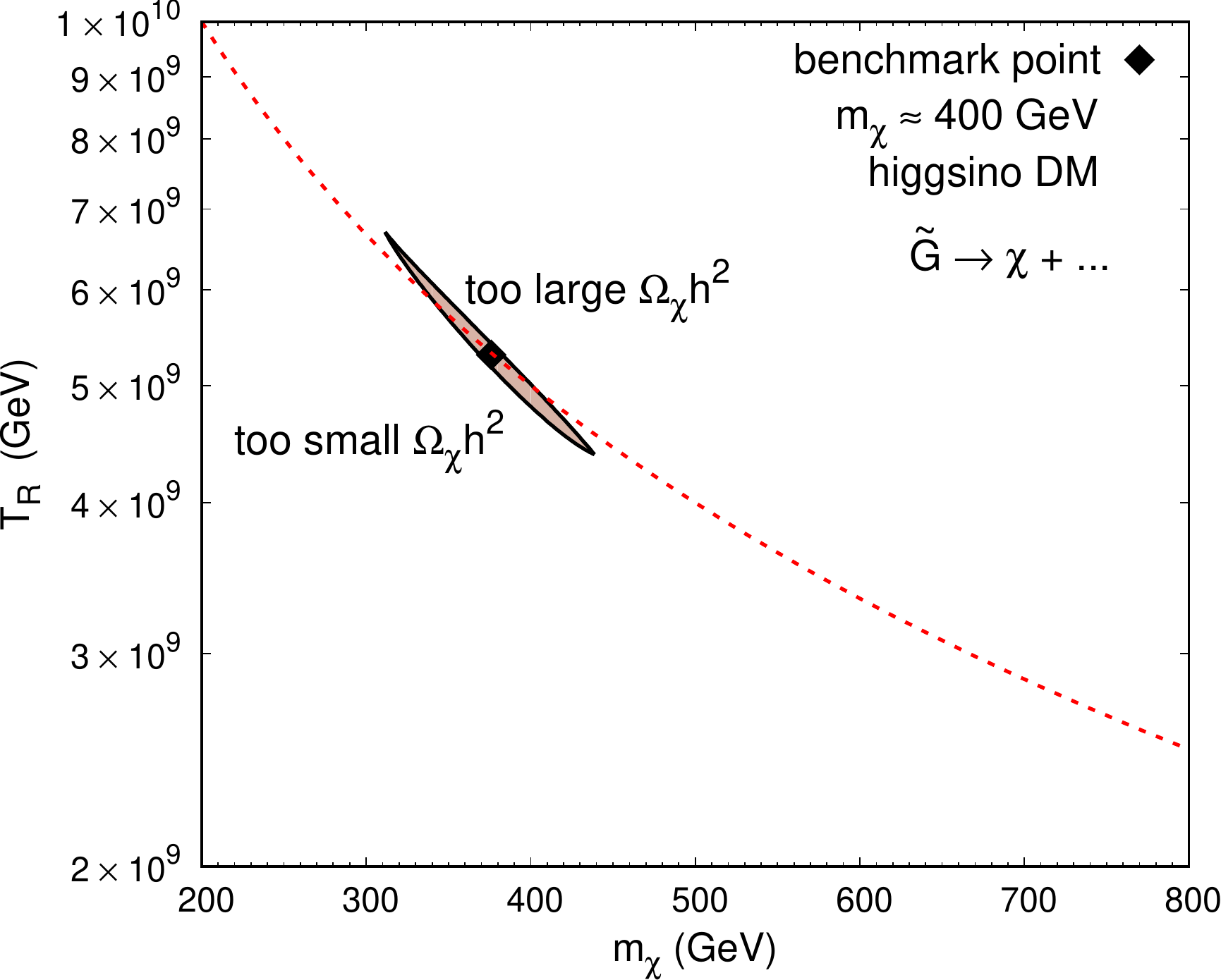}
\hfill
\includegraphics*[width=0.49\textwidth]{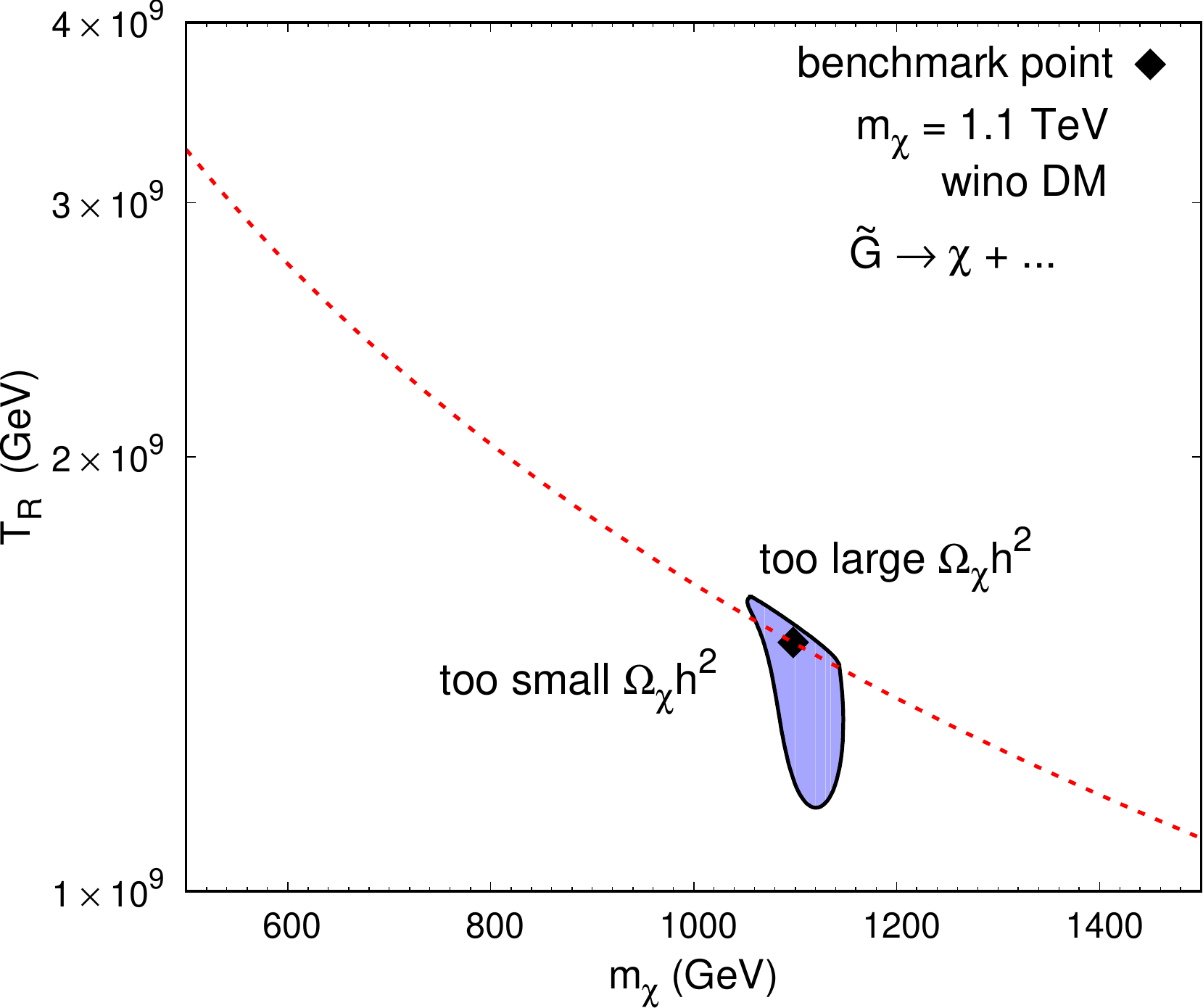}

\end{center}
\caption{The quality of reconstruction of the reheating temperature, $T_R$, as a function of the reconstructed mass of DM, $m_\chi$, for the same benchmark points as used in Fig.~\ref{fig:Oh2neutralino} for scenario with heavy gravitino ($m_{\widetilde{G}}\gtrsim 30$~TeV) decaying into higgsino (left panel) or wino (right panel) DM. 
Red dashed lines mark the lower bound on $T_R$ with only the particle physics constraints imposed assuming ideal reconstruction of $\Omega_\chi^{\textrm{non-th}} h^2$, \textsl{i.e.}, without the signal from a prospective DM
discovery taken into account.
} \label{fig:TRreconneutralino2}
\end{figure}

Gravitino interactions are suppressed by the Planck mass, $M_{\textrm{Pl}}\simeq 1.22\times 10^{19}$ GeV, which is several orders of magnitude larger than the Peccei-Quinn scale relevant for the axino, $f_a\simeq 10^9 - 10^{12}$ GeV \cite{Bae:2008ue}. As a result in order to obtain a correct amount of non-thermally produced neutralinos coming from late-time decays of gravitinos one needs to consider significantly larger reheating temperatures. 
Importantly, the  suppression by $M_{\textrm{Pl}}$ also results in longer lifetime of the gravitino that could lead to late-time decays violating the predictions of the BBN \cite{Kawasaki:2004qu,Jedamzik:2006xz}.
One also needs to take into account possible implications 
of the fact that the neutralinos produced in gravitino decays may form a warm fraction of neutralino DM.
In particular, it has been shown that this effect can become important for wino DM with mass around $100-500$ GeV \cite{Ibe:2012hr}.

Here 
we assume that the gravitino is significantly heavier than the neutralino, $m_{\widetilde{G}}\gg m_{\chi}$. More precisely, we assume that $m_{\widetilde{G}}\gtrsim 30$ TeV \cite{Kawasaki:2008qe}, which means that the gravitino is also significantly heavier than all the gauginos in our framework. The yield of thermally produced gravitinos is then practically independent of the gravitino mass.\
Similarly to the axino, the gravitino cannot be arbitrarily heavy in order to avoid a scenario in which it decays into neutralino before neutralino freezes out
and there is no non-thermal contribution to the neutralino DM density. 

The non-thermal contribution to the neutralino DM density in our scenario reads \cite{Bolz:2000fu,Pradler:2006qh,Rychkov:2007uq}
\begin{equation}
\Omega_{\chi}^{\textrm{non-th}}h^2\simeq 0.52\times \left(\frac{m_\chi}{1\,\textrm{TeV}}\right)\,\left(\frac{T_R}{10^{10}\,\textrm{GeV}}\right),\hspace{1.5cm}\textrm{$\chi$ from $\widetilde{G}$ decays}.
\label{eq:gravTP}
\end{equation}
Unlike in the axino case, for which
there was no such a simple estimate of $\Omega^{\textrm{non-th}}_\chi h^2$ like the one given in Eq.~(\ref{eq:gravTP}), because for low values $T_R$ the axino yield is not simply proportional to the reheating temperature and depends quite significantly on the MSSM spectrum \cite{Roszkowski:2015psa},
it is now straightforward to employ Eq.~(\ref{eq:gravTP}) to translate the limits on $\Omega_\chi^{\textrm{non-th}}h^2$ 
into constraints on $T_R$. We illustrate this in Fig.~\ref{fig:TRreconneutralino2} for both our benchmark scenarios. As expected the reheating temperature required to obtain the correct relic density is now significantly larger than for the axino. The quality of reconstruction is limited by both the uncertainty of $m_\chi$ and $\Omega_\chi^{\textrm{non-th}}h^2$, but in both scenarios $T_R$ can be determined with very good accuracy for wino DM benchmark scenario and up to a factor of $2$ for the higgsino DM case due to larger uncertainty on the DM mass
associated with 
the characteristic monochromatic-like feature present in annihilation spectrum for $W^+W^-$ final state that is more pronounced for larger DM masses.

\section{Conclusions}
\label{sec:conclusions}

In this paper, we addressed the question whether prospective detection of DM in the next generation of experiments could shed light on
whether DM was generated thermally in the freeze-out process in the early Universe. To this end, we simulated signals that could be seen in the
indirect detection experiments Fermi-LAT and CTA, as well as in the direct detection search in the Xenon1T experiment for DM particles with annihilation cross section significantly larger than required for thermal freeze-out. We then reconstructed the mass and the annihilation cross sections of such DM particles from these would-be signals and checked if the non-thermal component could be further understood in terms of appropriate underlying models.

We showed that in the model-independent approach the answer is negative except for a thin sliver in the parameter space assuming an $s$-wave annihilation. It could even become not possible once one considers, \textsl{e.g.}, the Sommerfeld enhancement of the annihilation cross section or multi-component DM scenarios. Hence some theoretical input is required for distinguishing the additional non-thermal component in the next decade or so. We discuss two such theoretical scenarios, varying in the degree of complexity: an EFT of DM with (axial) vector messenger and the MSSM. We showed that even with rather general assumptions about the EFT, the reconstruction of DM properties improves to the extent that the non-thermal component can be identified from the reconstruction of prospective ID and DD signals. We then turned to the MSSM and showed examples of benchmark points for which $m_\chi$ and/or $\langle \sigma v\rangle_0$ can be reconstructed even more precisely -- either because DM particle annihilate to $WW$ with important contribution from the Sommerfeld enhancement or because DM DD is sensitive to the gaugino composition of the neutralino. In addition, non-trivial correlations between the DD and ID rates, as well as the DM relic density allows to infer conclusions about the non-thermal contribution to the DM relic density in such a scenario even in the presence of effective coannihilations around the DM freeze-out. 

This enhanced precision of the determination of the non-thermal component can be used to study processes from which this component originated. We demonstrated this with two examples of EWIMPs constituting this component: the axino and the gravitino. In each case we determined the corresponding reheating temperature and concluded that -- compared to the present particle physics constraints -- future prospective detection of DM signal would lead to a moderate improvement of the allowed range of this parameter for the axino EWIMP, while for the gravitino EWIMP it would significantly narrow down the allowed range.

Taken together our results illustrate the possibility that the discovery of the DM signal alone -- though a qualitative step forward in characterization of the composition of the Universe -- is likely to give rise to a number of further questions. Answering these questions could require working in a specific theoretical framework, which would have to be inferred from other experiments. 

\acknowledgments
ST would like to thank A.~Hryczuk, T.M.P.~Tait and P.~Tanedo for helpful comments. 
LR is supported  in  part  by  the  National  Science  Council  (NCN)  research  grant  No.  2015-18-A-ST2-00748 and by the Lancaster-Manchester-Sheffield Consortium
for Fundamental Physics under STFC Grant No. ST/L000520/1.
ST is supported in part by the National Science Centre, Poland, under research
grant DEC-2014/13/N/ST2/02555, by the Polish Ministry of Science and Higher Education
under research grant 1309/MOB/IV/2015/0 and by NSF Grant No.~PHY-1620638. 
KT is partly supported by a grant 2014/14/E/ST9/00152 (National Science Centre, Poland).
The use of the CI\'S computer cluster at the National Centre for Nuclear Research is gratefully acknowledged.


\appendix

\section{Reconstruction of general DM particles}

\subsection{Gamma rays from DM annihilations}

Most of the $\gamma$ rays that reach detectors have an astrophysical origin which needs to be carefully taken into account when looking for signal from DM annihilation. However, one expects that DM-induced $\gamma$ rays are typically produced in dense regions with large $\rho_\chi$ and therefore they should manifest themselves as a small directional excess over the isotropic background. One such obvious region of interest for DM searches is the Galactic center (GC) where a significant increase of $\gamma$-ray flux from DM annihilations is expected due to the peak in the DM mass distribution. This is in particular true for the original NFW profile \cite{Navarro:1995iw}, as well as for the Einasto profile \cite{Merritt:2005xc,Graham:2006ae}, while
for the DM distributions with cores in the GC, \textsl{e.g.}, the Burkert profile \cite{Burkert:1995yz}, the increase of DM-induced $\gamma$-ray flux is not so pronounced.

Other promising sources of the $\gamma$ rays from DM annihilations are dSphs around the Milky Way. They are some of the most DM-dominated nearby objects known and therefore the DM searches focusing on dSphs suffer less from astrophysical background than in the case of the GC. However, the corresponding $J$ factors are typically smaller and their exact values are more difficult to determine. This introduce additional challenge when interpreting the experimental results, but still the most stringent limits on $\langle\sigma v\rangle_0$ up to date come from null DM searches in the combined analysis of FermiLAT and MAGIC observations of dSphs \cite{Ahnen:2016qkx,Fermi-LAT:2016uux}.
 
The total differential flux of the $\gamma$ rays from a source with the angular size $\Delta\Omega$ is given by
\begin{equation}
\frac{d\Phi}{dE}=\frac{\langle\sigma v\rangle_0}{8\pi m_\chi^2}\left(J_{\Delta\Omega}\sum_f{\textrm{BR}_f\frac{dN^f_{\gamma}}{dE}}
+\frac{1}{E^2}\int_{m_e}^{m_\chi}dE_s 
\bar{I}_{\textrm{IC},\Delta\Omega}(E,E_s)\frac{dN_{e^{\pm}}}{dE_s}\right)\,,\label{eq:DMflux}
\end{equation}
where $dN^f_{\gamma}/dE$ stands for the $\gamma$-ray spectrum for the annihilation final state $f$ with the corresponding branching ratio $\textrm{BR}_f$, the total annihilation cross section is given by $\langle\sigma v\rangle_0$, while the second term in the sum corresponds to the contribution from the Inverse Compton scatterings of the DM-induced electrons from the GC (see, \textsl{e.g.}, \cite{Cirelli:2010xx} for more details). In the case of the dSphs one does not expect significant contribution from the secondary $\gamma$-ray emission due to lower energy densities of photons than in the GC \cite{Lefranc:2016dgx}.

The astrophysical uncertainties connected to the DM distribution, $\rho$, are hidden in the $J$-factor which is defined as the integral of $\rho^2$ along the line of sight (l.o.s.) for the corresponding angular size of the DM source 
\begin{equation}
J_{\Delta\Omega}=\int_{\Delta\Omega} \int_{\textrm{l.o.s.}} 
\rho_\chi^2[r(\theta)]dr(\theta)d\Omega\,.
\label{eq:Jfac}
\end{equation}
In the case of the GC, we employ the generalized Navarro-Frenk-White (NFW) DM profile for $\rho_\chi$ that reads \cite{Zhao:1995cp}
\begin{equation}
\rho_\chi(r)=\frac{\rho_0\left(1+\frac{R_{\odot}}{r_s}\right)^{3-\gamma_{\textrm{NFW}}}}{\left(\frac{r}{R_{\odot}}\right)^{\gamma_{\textrm{NFW}}}\left(1+\frac{r}{r_s}\right)^{3-\gamma_{\textrm{NFW}}}},
\end{equation}
where we fix $r_s=20\textrm{ kpc}$ and the distance of the Solar System from the GC, 
$R_{\odot}=8.5\textrm{ kpc}$, while we let the remaining parameters, \textsl{i.e.}, $\rho_0$ and $\gamma_{\textrm{NFW}}$, to vary freely in the scan. For the dSphs we use the $J$ factors and their uncertainties from \cite{Ackermann:2013yva}.

\subsection{Direct detection of DM particles\label{Sec:DD}}

Another techniques of WIMP DM searches employs its possible scattering off heavy nuclei in deep underground detectors. The signal induced by DM particles can then be recorded as light coming from scintillation photons produced by de-exciting atoms, charge connected to atomic ionization or heat from phonons in crystals. In order to improve discrimination of the signal and background events coming from electron recoils one often employs two of the aforementioned detection strategies in a single experiment (for review see, \textsl{e.g.}, \cite{Undagoitia:2015gya}).

The distribution of the number of expected events, $R$, as a function of the recoil energy, $E_r$, of the DM particles off nuclei is given by
\begin{equation}
\frac{dR}{dE_r} = \frac{\sigma^{\textrm{SI}}_p}{m_\chi}\times\frac{A^2}{2\mu_p^2}\,F^2(E_r)\,\rho\, \int_{v\geq v_{\textrm{min}}}^{v\leq v_{\textrm{esc}}}{d^3v\,\frac{f(\mathbf{v},t)}{v}},
\label{eq:dRdE}
\end{equation}
where $\sigma_p^{\textrm{SI}}$ is the spin-independent WIMP-proton scattering cross section, $A$ is the atomic number of the nuclei, the reduced mass of the $\chi-p$ system is given by $\mu_p = m_pm_\chi/(m_p+m_\chi)$, $F(E_r)$ is the nuclear form factor, while the integral goes over the relative velocity between the WIMP DM particle and the nucleus, $v$, and the distribution of the WIMP velocities, $f(\mathbf{v},t)$, has a cut-off at the galaxy escape velocity, $v_{\textrm{esc}}$. The minimum velocity that can result in an event with the recoil energy $E_r$ is given by $v_{\textrm{min}} = \sqrt{M\,E_r/(2\,\mu^2)}$, where $M$ is the mass of the nucleus and $\mu = m_\chi M/(m\chi + M)$ is the reduced mass of the $\chi$-nucleus system. We assume that the DM particles scatter elastically off nuclei and that there are no important isospin violation effects. Inelastic scatterings are typically suppressed with respect to the elastic ones \cite{Ellis:1988nb,Engel:1999kv} and we neglect such possibility in our study (for reconstruction with inelastic scatterings and/or isospin violation effects taken into account see, \textsl{e.g.}, \cite{Pato:2011de,Newstead:2013pea,Peter:2013aha}).

If the WIMP DM mass is larger than the nucleus mass, the minimal velocity, $v_{\textrm{min}}$, is insensitive to $m_\chi$ and therefore the recoil energy distribution is simply proportional to $\sigma_p^{\textrm{SI}}/m_\chi$ as shown in Eq.~(\ref{eq:dRdE}). This makes it impossible to reconstruct the properties of such heavy DM particles from only DD experiments at a level of model-independent WIMP DM since only the ratio $\sigma_p^{\textrm{SI}}/m_\chi$ can be inferred from the data. However, the interplay between DD and ID can help to overcome this problem \cite{Bernal:2008zk,Arina:2013jya,Kavanagh:2014rya,Roszkowski:2016bhs}, especially in a framework of particular models in which one can benefit from the cross correlation between $\sigma_p^{\textrm{SI}}$ and $\langle\sigma v\rangle_0$.

\subsection{General methodology of reconstruction\label{sec:methodology}}

In order to determine the quality of reconstruction of the DM properties, we first generate a signal mock data set for would-be discovered DM particles defined by our assumed benchmark points. The reconstruction is then performed by scanning over the parameters of the models that we consider (see Tables~\ref{tab:modelindepparams}, \ref{Tab:paramsEFT} and \ref{tabp10MSSM}), as well as over the nuisance parameters that define the astrophysical uncertainties (see Table~\ref{tab:nuiparams}). For each point in the scan we estimate the quality of fit of the obtained data set to the fixed signal mock data set of the benchmark point by evaluating the respective likelihood functions. We finally present our results in 2-dimensional plots showing $95\%$ confidence level (CL) regions defined by the condition $\Delta\chi^2 = 5.99$.

\begin{table}[t]
   \centering\footnotesize
   \begin{tabular}{|c|c|c|c|} 
      \hline
      \textbf{Symbol} & \textbf{Parameter} & \textbf{Scan range} & \textbf{Prior distribution} \\
      \hline
      \hline
       $v_0$ & Circular velocity & $220\pm 20$ km/s & Gaussian \\
       \hline
       $v_{\textrm{esc}}$ & Escape velocity & $544\pm 40$ km/s & Gaussian  \\
       \hline
       $\rho_0$ & Local DM density & $0.3\pm 0.1\,\textrm{GeV}/\textrm{cm}^3$ & Gaussian \\
       \hline
       $\gamma_{\textrm{NFW}}$ & NFW slope parameter & $1.20\pm 0.15$ & Gaussian \\
       \hline   
   \end{tabular}
   
   \caption{Nuisance parameters that define the astrophysical uncertainties in our scans.}
   \label{tab:nuiparams}
\end{table}

The details of the likelihood functions and the reconstruction methodology can be found in \cite{Roszkowski:2016bhs}. In our reconstruction we base on the code developed for that study. Here we only briefly summarize the main aspects of the procedure. 
In the case of the GC we analyze $\gamma$-ray signals from DM annihilations that will be observed by the CTA for which we assume $500$~hours of observational time. We take into account the background from the Galactic Diffuse Emission (GDE) and cosmic rays (CRs). In the fitting, we use the binned Poisson likelihood function convoluted with the Gaussian functions describing the uncertainties of the background estimation, where the bins are defined both in the photon energy, as well as for the four spatial regions in the sky around the GC.  
In the study of the DM-induced $\gamma$-ray signal from the dSphs we focus on DM searches performed by the FermiLAT. For this analysis, we assume $15$ years of observation and an extended set of $46$ dSphs. The likelihood function in the fitting procedure follows \cite{Ackermann:2013yva} from where we also take the uncertainties on the $J$ factors.

In the case of the DM DD we employ a binned Poisson likelihood in which the expected signal is obtained by the integration of Eq.~(\ref{eq:dRdE}) in each energy bin. The residual background in DD experiments is typically small due to a high purity of target material and good separation from external sources of radioactivity. In addition, one of the main sources of the background, namely the electronic recoils, can be discriminated from the nuclear recoils thanks to a combination of two detection strategies, as in the dual-phase Xenon1T time projection chamber that records signal both as scintillation light and ionization charge. 


\section{Reconstruction of neutralino DM particles\label{App:B}}

\subsection{Direct detection and gamma rays for indirect detection of neutralino DM}

Higgsino and wino DM particles annihilate dominantly into $W^+W^-$, $ZZ$, $Zh$ and $q\bar{q}$ final states with possible admixtures of other channels that involve non-SM particles, \textsl{e.g.}, $W^\pm H^\pm$ and $hA$. This leads to a continuous spectrum of photons originating from subsequently produced cascades of particles. Additional contributions to $\gamma$-ray spectrum comes from internal bremsstrahlung processes, as well as, more importantly, from loop-suppressed, but non-negligible annihilations into $\gamma\gamma$ and $Z\gamma$ pairs that result in monochromatic lines. For completeness, we also take into account secondary emission of $\gamma$ rays from the inverse Compton scattering of DM-induced electrons, although it typically plays a subdominant role for higgsino and wino DM due to suppressed annihilation branching ratios into leptons. When generating the annihilation spectra we employ tables provided in \cite{Cirelli:2010xx,Ciafaloni:2010ti}, with the additional contribution from secondary $\gamma$ rays that follows \cite{Buch:2015iya}. To obtain the cross sections for monochromatic lines, as well as $\gamma$-ray spectrum from internal bremmstrahlung processes we employ MicrOMEGAs 4.3.1 \cite{Belanger:2014vza}. In the case of the annihilation channels with products from beyond the SM, \textsl{e.g.}, $hA$ or $W^\pm H^\pm$, which are not treated in \cite{Cirelli:2010xx}, we employ SUSY-HIT \cite{Djouadi:2006bz} to determine their branching ratios into the SM particles.
We then use for these SM decay products the $\gamma$-ray spectra from \cite{Cirelli:2010xx} obtained for the appropriately shifted CM energies. Such a simplified approach is the most relevant for the annihilation products 
produced at rest, which we find applicable to the points in our scan that have significant branching ratios into the non-SM particles. In other scenarios the approach is based on the assumption that the \textsl{a 
priori} smooth distribution of energies of final SM particles in the CM frame of the annihilating DM particles can be approximated by the average values of these energies. We would like to stress here that this approximation does 
not affect much our final results since most of the points have dominant 
annihilation final states into the SM particles. In the case of wino DM, one also needs to take into account the Sommerfeld enhancement which can significantly modify the annihilation cross section both when calculating the DM freeze-out density and $\langle\sigma v\rangle_0$ \cite{Hisano:2004ds,Hisano:2006nn} (see also \cite{Hryczuk:2011vi,Fan:2013faa,Cohen:2013ama,Hryczuk:2014hpa,Beneke:2016ync,Ovanesyan:2016vkk} for more recent studies). Here we follow \cite{Hryczuk:2011tq,Ovanesyan:2014fwa}.

The dominant contribution to the spin-independent cross section, $\sigma_p^{\textrm{SI}}$, for higgsino and wino DM comes from a $t$-channel Higgs boson(s) exchange unless $\chi$ is a pure state. In the latter case both the $s$-channel squark exchange, whose rate is suppressed by the squark mass, as well as loop-induced processes become more important \cite{Hisano:2004pv}, but the scattering cross section that one obtains in such a scenario typically lies below the experimentally accessible limits. In accord with our approach of focusing on the scenarios that can be probed by DM DD in the following years,
we choose our benchmark points to have subdominant but non-negligible mixing between gauginos and higgsino.


\subsection{Methodology of reconstruction for neutralino DM\label{sec:reconneutralino}}

We generate the signal mock data set for both benchmark points and subsequently attempt to fit it with the signal corresponding to other neutralino DM scenarios obtained in the scan over the parameter space of the MSSM. We perform the scan over a 10-parameter version of the MSSM with the parameters and their ranges  shown in Table~\ref{tabp10MSSM}. We employ Mutlinest \cite{Feroz:2009} for sampling the parameter space of the model. We use SOFTSUSY-3.4.0 \cite{Allanach:2002} to generate the mass spectrum and we take into account the requirement that the mass of the lightest SM-like Higgs boson, $m_h$, fits the observed value \cite{Aad:2015zhl} with $3$ GeV theoretical error. We also take into account the LHC limits on gluino and squark masses following \cite{Kowalska:2016ent}, as well as basic constraints related to $B$-physics, for which we use SuperIso v3.3 \cite{Arbey:2009}, although they play minor  role in our discussion. The constraints imposed in the scan other than fitting to the signal mock data set are summarized in Table~\ref{tab:constraintsMSSM}.

\begin{table}[t]
\centering
\begin{tabular}{|c|c|}
\hline 
Parameter & Range \\ 
\hline
\hline 
bino mass & $0.1 < M_1 < 5$ \\ 
wino mass & $0.1 < M_2 < 6$ \\ 
gluino mass & $0.7 < M_3 < 10$ \\ 
stop trilinear coupling & $-12 < A_t < 12$ \\ 
stau trilinear coupling & $-12 < A_{\tau} < 12$ \\ 
sbottom trilinear coupling & $A_b = -0.5$ \\ 
pseudoscalar mass & $0.2 < m_A < 10$ \\ 
$\mu$ parameter & $0.1 < \mu < 6$ \\ 
3rd gen. soft squark mass & $0.1 < m_{\widetilde{Q}_3} < 15$ \\ 
3rd gen. soft slepton mass & $0.1 < m_{\widetilde{L}_3} < 15$ \\ 
1st/2nd gen. soft squark mass  & $m_{\widetilde{Q}_{1,2}} = M_1 + 1$ TeV \\ 
1st/2nd gen. soft slepton mass  & $m_{\widetilde{L}_{1,2}} = m_{\widetilde{Q}_3} + 100$ GeV \\ 
ratio of Higgs doublet VEVs & $2 < \tan\beta < 62$ \\ 
\hline 
\hline
Nuisance parameter & Central value, error \\
\hline
\hline
Bottom mass $m_b(m_b)^{\bar{MS}}$ (GeV) & (4.18, 0.03) \cite{Olive:2016xmw}\\
Top pole mass $m_t$ (GeV) & (173.21, 0.87) \cite{Olive:2016xmw}\\
\hline
\end{tabular}
\caption{The parameters of the p10MSSM and their
  ranges used in our scan. All masses and trilinear couplings are given in TeV, unless
  indicated otherwise. All the parameters of the model are given at the
  SUSY breaking scale.} 
\label{tabp10MSSM} 
\end{table}

\begin{table}[H]
\centering
\begin{tabular}{|c|c|c|c|}
\hline 
Measurement & Mean & Error: exp., theor. & Ref. \\ 
\hline
\hline 
$m_h$ & $125.09$ GeV  & $0.24$ GeV, $3$ GeV & \cite{Aad:2015zhl}\\ 
BR($\widebar{\textrm{B}}\rightarrow X_s\gamma$)$\times 10^4$ & $3.43$ & $0.22$, $0.21$ & \cite{Amhis:2014hma,SLAC} \\ 
BR($\textrm{B}_{\textrm{u}}\rightarrow \tau\nu$)$\times 10^4$ & $1.09$ & $0.24$, $0.38$ & \cite{Olive:2016xmw}\\ 
$\Delta M_{B_s}$ & $17.756$ $\textrm{ps}^{-1}$ & $0.021\,\textrm{ps}^{-1}$, $2.400\,\textrm{ps}^{-1}$ & \cite{Olive:2016xmw}\\
$\sin^2\theta_{\textrm{eff}}$ & $0.23155$& $0.00005$, $0.00015$ & \cite{Olive:2016xmw}\\
$M_W$ & $80.385$ GeV & $0.015$ GeV, $0.015$ GeV & \cite{Olive:2016xmw}\\
BR($\textrm{B}_{\textrm{s}}\rightarrow\mu^+\mu^-$)$\times 10^9$ & $2.9$ & $0.7$, $10\%$ & \cite{Aaltonen:2013as,CMS:2014xfa}\\
\hline 
\end{tabular}
\caption{The constraints imposed on the parameter spaces of the p10MSSM other than fitting to the DM signal mock data set
and LHC SUSY searches. 
}
\label{tab:constraintsMSSM} 
\end{table}


\end{document}